\documentclass{article}
\usepackage{iclr2026_conference,times}

\usepackage{hyperref}
\usepackage{url}
\usepackage{graphicx}
\usepackage{booktabs}
\usepackage{amsmath}
\usepackage{multirow}
\usepackage{placeins}
\usepackage{float}
\usepackage{subcaption}
\usepackage{enumitem}

\usepackage{xurl}      
\usepackage{hyperref}  
\usepackage{xcolor}

\newcommand{\keytakeaway}[1]{%
  \begingroup
  \setlength{\fboxsep}{0pt}%
  \colorbox{blue!20}{#1}%
  \endgroup
}

\title{Diagnosing FP4 Inference: A Layer-Wise and Block-Wise Sensitivity Analysis of NVFP4 and MXFP4}

\author{Musa Cim \\
Department of Computer Science and Engineering \\
The Pennsylvania State University \\
University Park, PA, USA \\
\texttt{mtc5693@psu.edu}
\And
Burak Topcu \\
Department of Computer Science and Engineering \\
The Pennsylvania State University \\
University Park, PA, USA \\
\texttt{bvt5283@psu.edu}
\And
Mahmut Taylan Kandemir \\
Department of Computer Science and Engineering \\
The Pennsylvania State University \\
University Park, PA, USA \\
\texttt{mtk2@psu.edu}
}

\iclrfinalcopy
\begin{document}
\raggedbottom

\maketitle

\begin{abstract}
    Quantization addresses the high resource demand for large language models (LLMs) by alleviating memory pressure and bandwidth congestion and providing significantly scaled compute power with a tolerable impact on accuracy. Four-bit floating point (FP4), the lowest-precision format that preserves essential numerical properties such as exponent and sign, has begun to be adopted in cutting-edge architectures, including Blackwell and AMD CDNA, to support LLM quantization and reduce deployment costs. Although aggressive quantization can yield efficiency gains, the quantization sensitivity of within-transformer layers and whether these sensitivities generalize across existing FP4 formats and model scales remain underexplored. To elucidate quantization sensitivity, this study conducts a systematic analysis of two FP4 formats, MXFP4 and NVFP4, across three Qwen2.5 model scales (0.5B, 7B, and 14B), using controlled component-wise and block-wise isolation methodologies. We observe that MLP up- and down-projection layers consistently dominate in terms of sensitivity, while gate and attention projections are moderately and substantially less sensitive to FP4 quantization, respectively. We further find that sensitivity does not universally localize to the final blocks, but early blocks can be highly sensitive, particularly under MXFP4. Our results provide a diagnostic characterization of the inference behavior of FP4 across components, depths, and FP4 formats.
\end{abstract}

\section{Introduction} \vspace{-2.5mm}
During the past decade, large language models (LLMs) have reshaped both the AI research paradigm and industrial practices~\citep{AiIndexRep}, and empirical scaling laws~\citep{scalingLaw} demonstrate that scaling up the size of the model can lead to further performance improvements. However, deploying larger-scale models--trained over prolonged runs across thousands of server-class accelerators--substantially increases the costs and energy consumption required to power LLM-based applications. Quantization~\citep{quant1,quant2}, which narrows the numerical precision, has emerged as a promising technique to alleviate pressure on memory and bandwidth, scale the logical computational capability (e.g., TFLOPs), and thus reduce the cost of model deployment. Compared with traditional full- or half-precision formats (e.g., 32-bit and 16-bit), modern accelerators now support lower-precision representations, including 8-bit~\cite{quant2}, 6-bit, and even 4-bit formats~\citep{nvidia2024nvfp4, amd2024mxfp4}, enabling the practical deployment of aggressive quantization techniques.

In particular, FP4 quantization has seen a growing adoption, with formats such as MXFP4~\citep{amd2024mxfp4} and NVFP4~\citep{nvidia2024nvfp4} increasingly used in both the development and deployment of LLMs~\citep{LLM_FP4, nvfp4_pretraining, mlperf_fp4_2025}. Prior work has established that quantization error in transformers is highly non-uniform across layers and components~\citep{fan2019reducing,frantar2022gptq,lin2024awq}, and that rare activation outliers dominate low-bit quantization error~\citep{dettmers2022gpt3,xiao2023smoothquant}, motivating component-aware handling. Additionally, previous studies on layer-wise importance~\citep{xiong2020layer,zhang2024investigating,nepal2025layer,skean2502layer} often emphasize that later layers play a dominant role in shaping model outputs and task performance. However, to our knowledge, no comprehensive component- and block-wise sensitivity analysis exists specifically for FP4 formats, leaving open questions about which architectural elements are most affected and how sensitivity varies across model scales. To address this gap, we analyze the quantization sensitivity of the MXFP4 and NVFP4 formats across three Qwen2.5 model scales (0.5B, 7B, and 14B), using controlled component- and block-wise isolation methodologies.

This paper makes the following {\bf contributions}:

\begin{itemize}[leftmargin=2.5em, itemsep=1pt, topsep=1pt, parsep=0pt, partopsep=0pt]
    \item It presents an analysis indicating that up- and down-projections in the MLP layer consistently form the most sensitive tier, and this trend remains stable across FP4 formats and model scales. 
    \item It observes, across blocks, that the FP4 sensitivity is not necessarily confined to the final blocks; for the 0.5B model under MXFP4, earlier blocks exhibit strong sensitivity, challenging the common assumption that the last blocks dominate.
    \item It reveals that extreme activation outliers are consistent with the high sensitivity of the down projection, but do not fully account for FP4 sensitivity, as the up projection is comparably sensitive despite lower outlier ratios. 
    \item It demonstrates that the model scale affects the magnitude of sensitivity but not the relative sensitivity between components.
\end{itemize}

\section{Background and Motivation} \vspace{-2.5mm}
\noindent {\bf Background:} Large language models (LLMs) impose substantial memory bandwidth and compute demands, making low-precision inference a central technique for efficient deployment. Quantization formats such as FP16 and FP8 have demonstrated strong accuracy and efficiency tradeoffs~\citep{mixedPrec, fp8DL} and are now widely supported in production inference and training pipelines~\cite{mlperf_main}. More recently, proposals have emerged for inference in ultra-low-precision regimes, particularly with variants of FP4 that preserve sign and exponent while further reducing storage and bandwidth demands~\cite{}, and accelerator vendors such as NVIDIA and AMD have introduced both native FP4 tensor-operation support and specialized FP4 formats (e.g., NVFP4~\citep{nvidia2024nvfp4} and MXFP4~\citep{amd2024mxfp4}), along with hardware-accelerated scaling mechanisms, making FP4 inference increasingly practical rather than purely algorithmic~\citep{amd2024precision}. As a result, FP4 has transitioned from a research concept to a deployable precision target for large-scale LLM inference.

\noindent {\bf Motivation:} Although FP4 inference is now supported by modern accelerators, it is increasingly evident that there is no universal recipe for applying FP4 across different models, formats, and applications, as aggressive quantization can significantly degrade inference quality. Prior work in low-bit quantization has shown that quantization behavior is highly sensitive to activation distributions and data characteristics, motivating activation-aware and data-dependent strategies rather than uniform precision assignments~\citep{lin2024awq,xiao2023smoothquant}. In parallel, different FP4 formats employ distinct scaling mechanisms and calibration assumptions, leading to format-specific considerations even for the same model architecture~\citep{abecassis2025pretraining}. Hence, FP4 sensitivity is jointly influenced by model architecture, format preference, and data distribution, and heuristics that are effective in one setting may fail in another. Consequently, a principled FP4 deployment requires a detailed analysis of FP4 sensitivity within transformer layers and across blocks to avoid accuracy degradation. This motivates a controlled component- and block-wise sensitivity analysis to diagnose FP4 failure modes and to provide a foundation for adapting FP4 recipes to specific models, formats, and deployment scenarios.

\section{Experimental Design} \vspace{-2.5mm}
We adopt a controlled isolation methodology that quantizes individual components and blocks while keeping all other factors fixed. Under this setting, we formulate four hypotheses:

\begin{enumerate} [leftmargin=2.5em, itemsep=1pt, topsep=1pt, parsep=0pt, partopsep=0pt]
    \item \textbf{Mechanism:} Component sensitivity to FP4 is primarily determined by activation tail-heaviness (e.g., Max/P99) rather than by whether the component belongs to MLP or attention.
    \item \textbf{Isolation/Ordering:} When components are isolated and quantized one at a time, their sensitivity ranking remains consistent across experiments.
    \item \textbf{Model Scale:} Increasing model size (0.5B $\rightarrow$ 7B $\rightarrow$ 14B) mainly affects the magnitude of sensitivity, not the relative ordering of the sensitive components. 
    \item \textbf{Block Localization:} FP4 sensitivity does not always peak in the final blocks; early-block sensitivity can emerge, particularly under MXFP4.
\end{enumerate}

We experiment with three different scales of Qwen2.5 models: 0.5B (24 layers), 7B (28 layers), and 14B (48 layers) to assess scale generalization. Across FP4 formats, we compare MXFP4 (E2M1, 32-element blocks, shared 8-bit exponent~\citep{mxfp4_exp}) and NVFP4 (dynamic scaling, 16-element blocks, 4-bit scales, max calibration algorithm~\citep{nvidia2024nvfp4}). Experiments are conducted on RTX 5090~\citep{rtx5090} for smaller models and RTX 6000 Pro~\citep{rtxpro6000} for larger models. Perplexity is measured on WikiText-2~\citep{wikitext} using 256 calibration samples.

We use on-the-fly quantization such that weights are stored in FP16 and quantized to FP4 during inference, then dequantized back for computation. For component sensitivity, we quantize one of all seven projection types (Query, Key, Value, Output, Gate, Up, Down) to FP4 at a time, and keep remaining six of them in FP16 across all layers. For block sensitivity, we keep one specific block's component in FP16 while the remaining six component types plus other blocks of the same type are quantized to FP4.

\begin{figure}[H]
\centering
\begin{subfigure}[b]{0.24\textwidth}
    \centering
    \includegraphics[width=\textwidth]{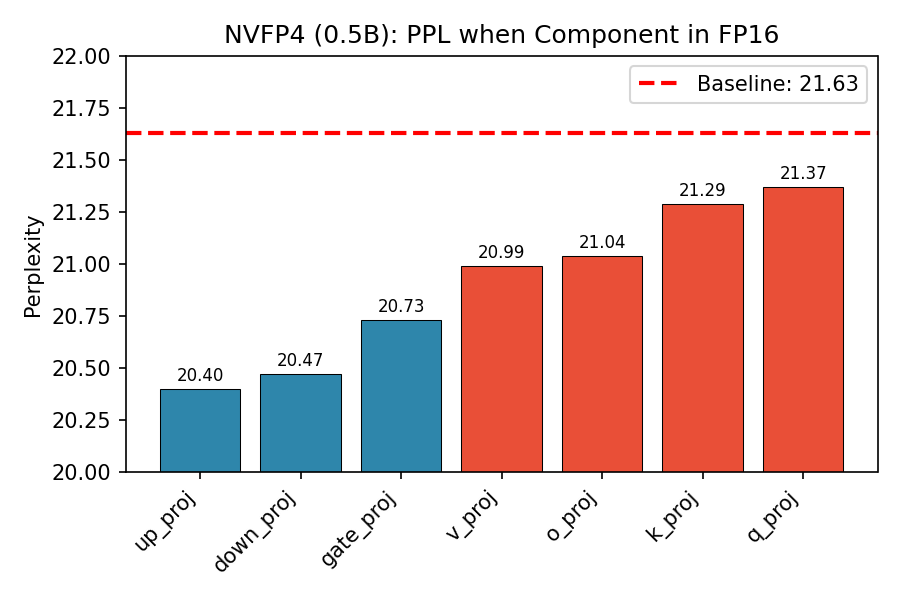}
    \caption{NVFP4 0.5B PPL}
\end{subfigure}
\hfill
\begin{subfigure}[b]{0.24\textwidth}
    \centering
    \includegraphics[width=\textwidth]{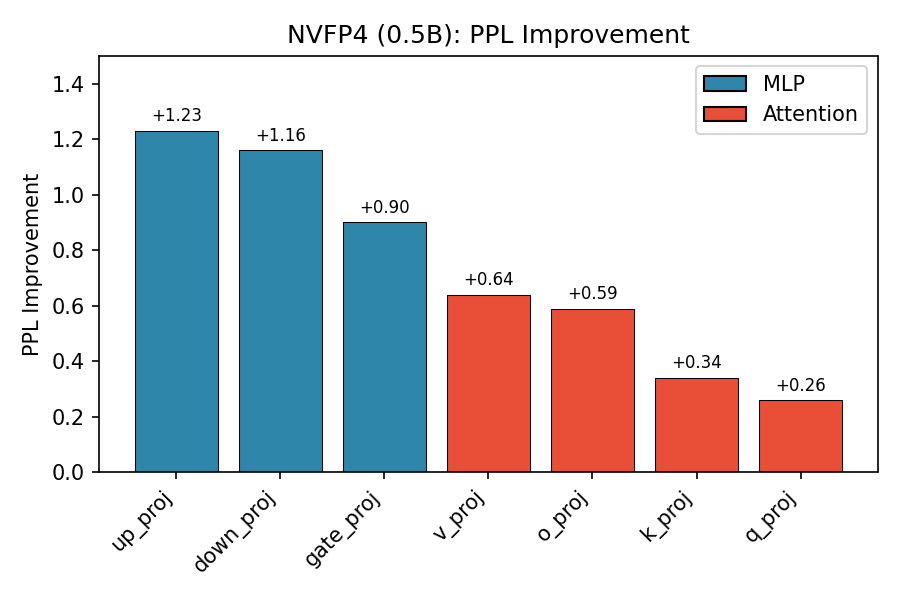}
    \caption{NVFP4 0.5B Improv.}
\end{subfigure}
\hfill
\begin{subfigure}[b]{0.24\textwidth}
    \centering
    \includegraphics[width=\textwidth]{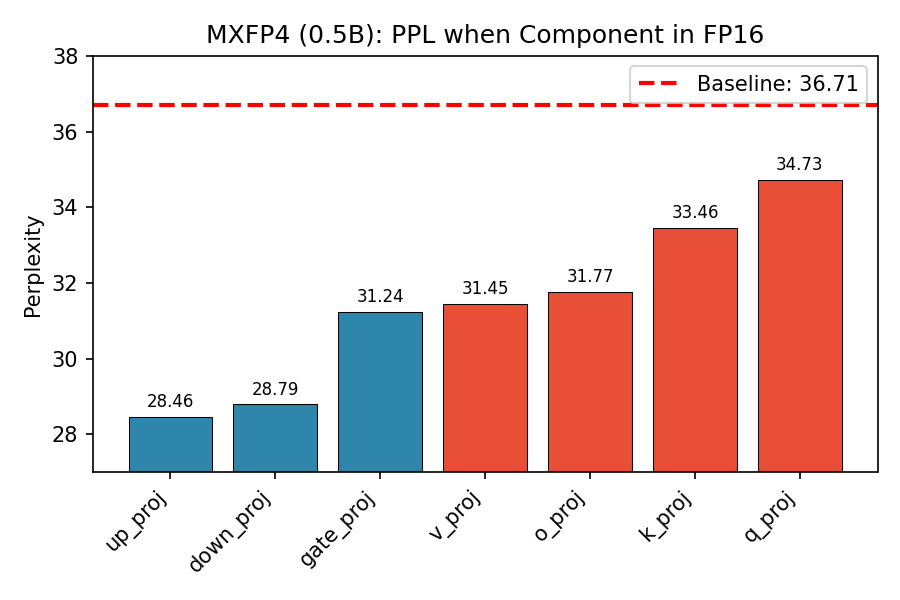}
    \caption{MXFP4 0.5B PPL}
\end{subfigure}
\hfill
\begin{subfigure}[b]{0.24\textwidth}
    \centering
    \includegraphics[width=\textwidth]{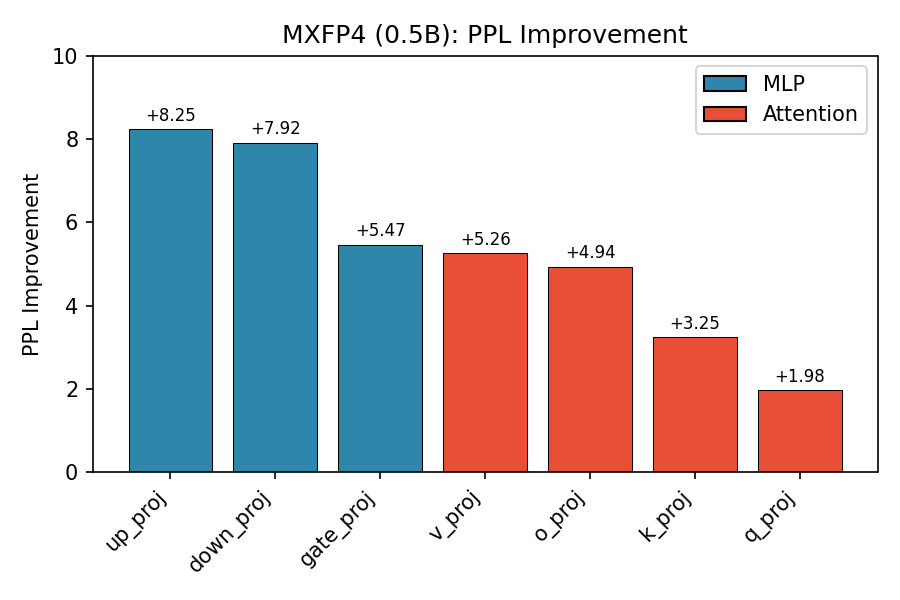}
    \caption{MXFP4 0.5B Improv.}
\end{subfigure}

\vspace{0.15cm}

\begin{subfigure}[b]{0.24\textwidth}
    \centering
    \includegraphics[width=\textwidth]{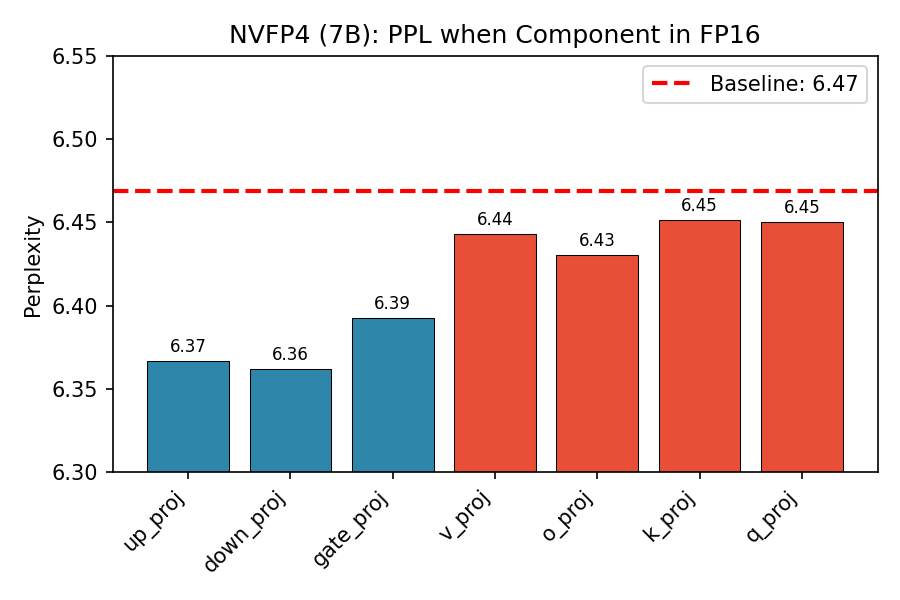}
    \caption{NVFP4 7B PPL}
\end{subfigure}
\hfill
\begin{subfigure}[b]{0.24\textwidth}
    \centering
    \includegraphics[width=\textwidth]{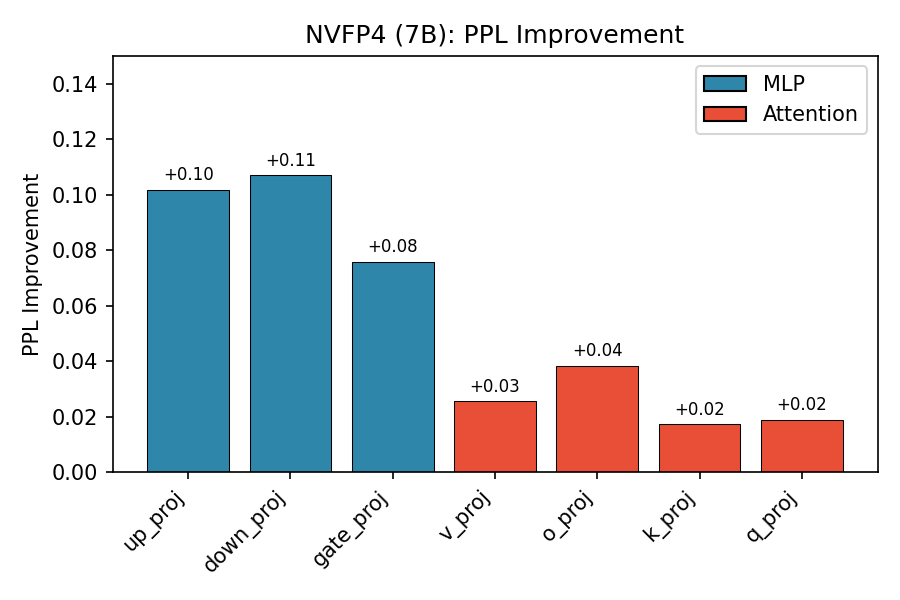}
    \caption{NVFP4 7B Improv.}
\end{subfigure}
\hfill
\begin{subfigure}[b]{0.24\textwidth}
    \centering
    \includegraphics[width=\textwidth]{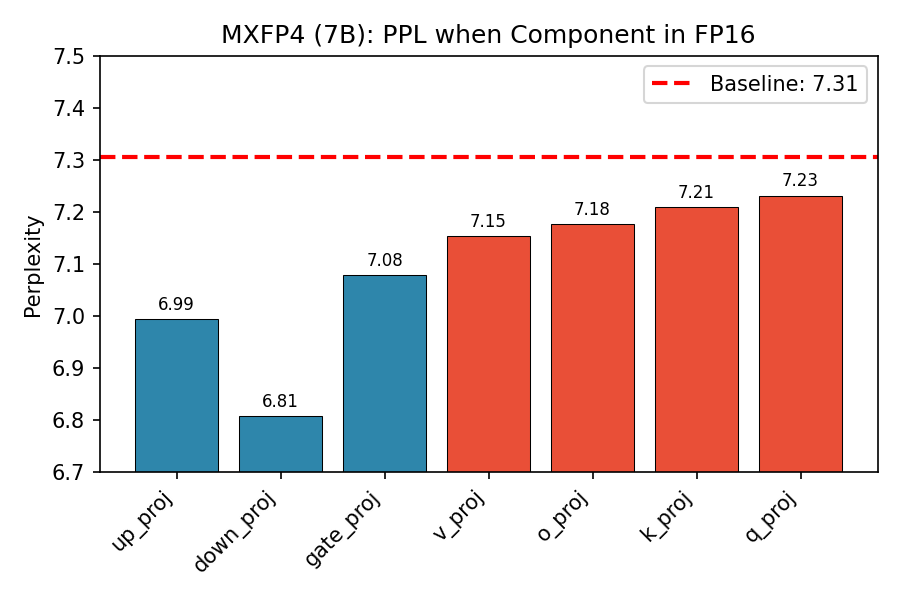}
    \caption{MXFP4 7B PPL}
\end{subfigure}
\hfill
\begin{subfigure}[b]{0.24\textwidth}
    \centering
    \includegraphics[width=\textwidth]{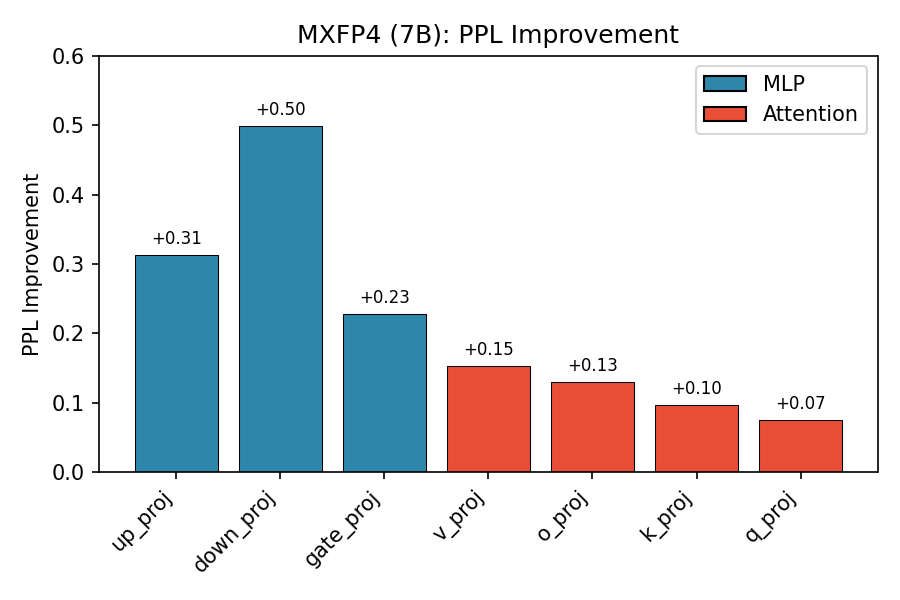}
    \caption{MXFP4 7B Improv.}
\end{subfigure}

\vspace{0.15cm}

\begin{subfigure}[b]{0.24\textwidth}
    \centering
    \includegraphics[width=\textwidth]{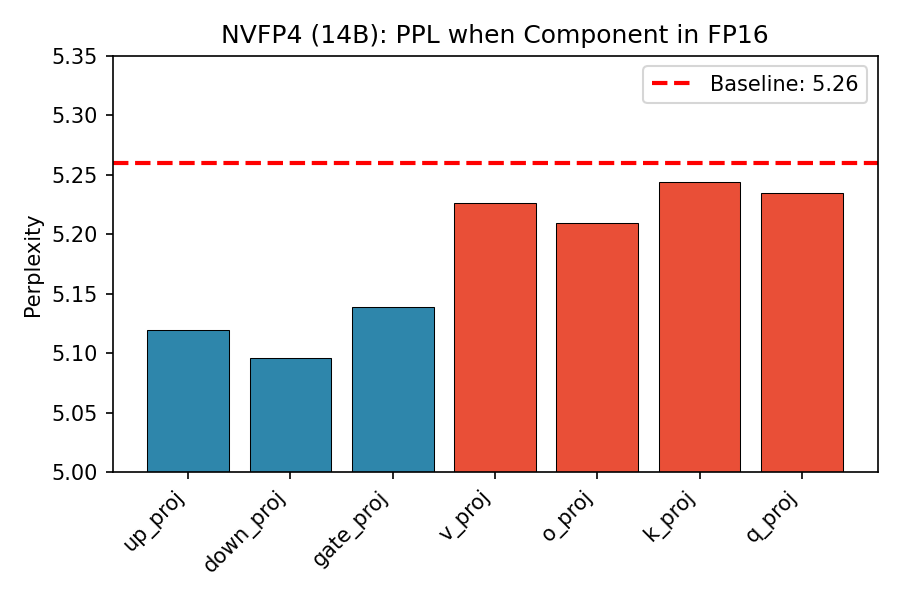}
    \caption{NVFP4 14B PPL}
\end{subfigure}
\hfill
\begin{subfigure}[b]{0.24\textwidth}
    \centering
    \includegraphics[width=\textwidth]{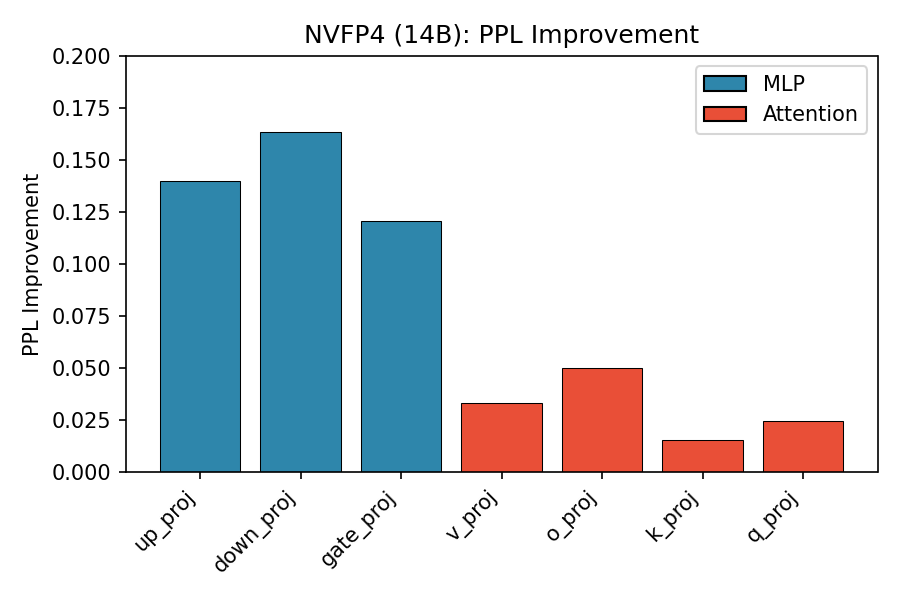}
    \caption{NVFP4 14B Improv.}
\end{subfigure}
\hfill
\begin{subfigure}[b]{0.24\textwidth}
    \centering
    \includegraphics[width=\textwidth]{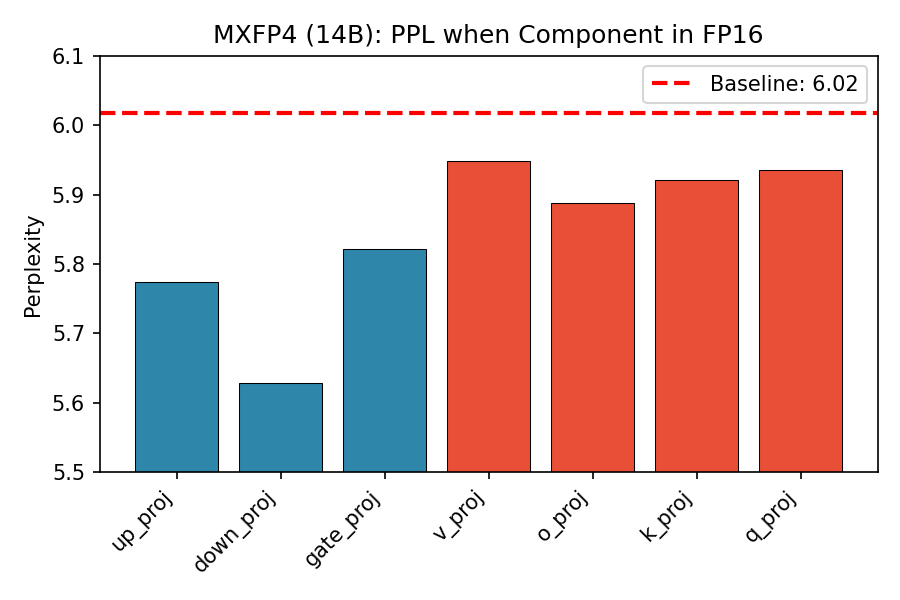}
    \caption{MXFP4 14B PPL}
\end{subfigure}
\hfill
\begin{subfigure}[b]{0.24\textwidth}
    \centering
    \includegraphics[width=\textwidth]{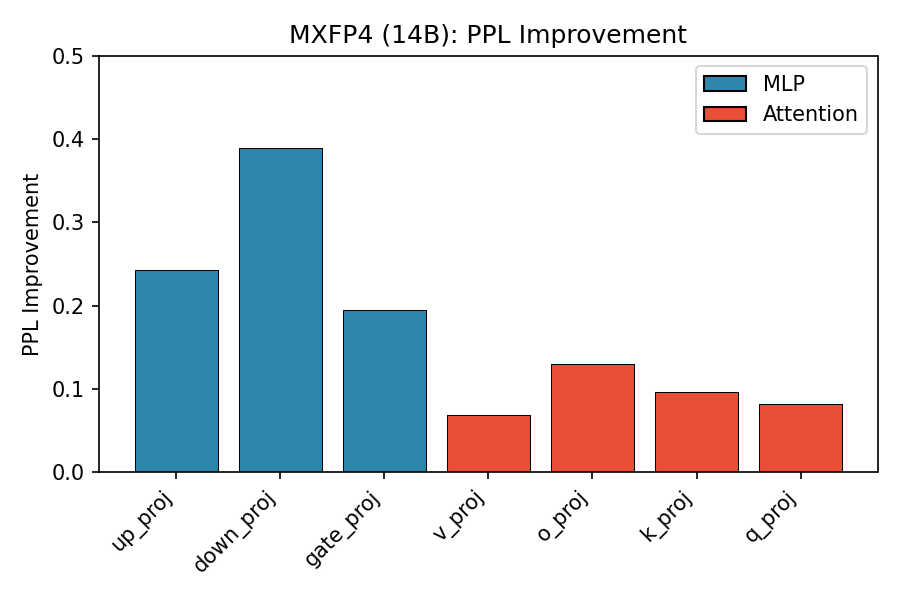}
    \caption{MXFP4 14B Improv.}
\end{subfigure}
\vspace{-1mm}
\caption{Component sensitivity comparison across three model scales. Rows: 0.5B, 7B, 14B. Blue = MLP, Red = Attention. MLP projections (down and up) consistently form the most sensitive tier across all scales and formats.}
\label{fig:component}
\vspace{-4mm}
\end{figure}

\section{Results} \vspace{-2.5mm}

\noindent {\bf Component Sensitivity.} Figure~\ref{fig:component} shows that FP4 sensitivity is highly non-uniform across transformer components. Across all model scales and both FP4 formats, the quantization sensitivity is dominated by MLP projections, with up- and down-projections consistently forming the most sensitive tier, while gate and attention projections are moderately and substantially less sensitive, respectively. Although larger models become more sensitive to quantization—and MXFP4 shows higher sensitivity than NVFP4, the relative tiering of components remains stable across all settings. Tables~\ref{tab:component}, ~\ref{tab:component_7b}, and \ref{tab:component_14b} in the Appendix provide detailed numerical results regarding the sensitivity for 0.5B, 7B, and 14B model scales, respectively.

\keytakeaway{\textbf{Key Takeaway \#1:}} MLP projections are highly sensitive and consistently behave the most sensitive tier to FP4 quantization, whereas gate and attention projections are moderately and substantially less sensitive, respectively. 

\begin{figure}[H]
\centering
\includegraphics[width=0.95\linewidth]{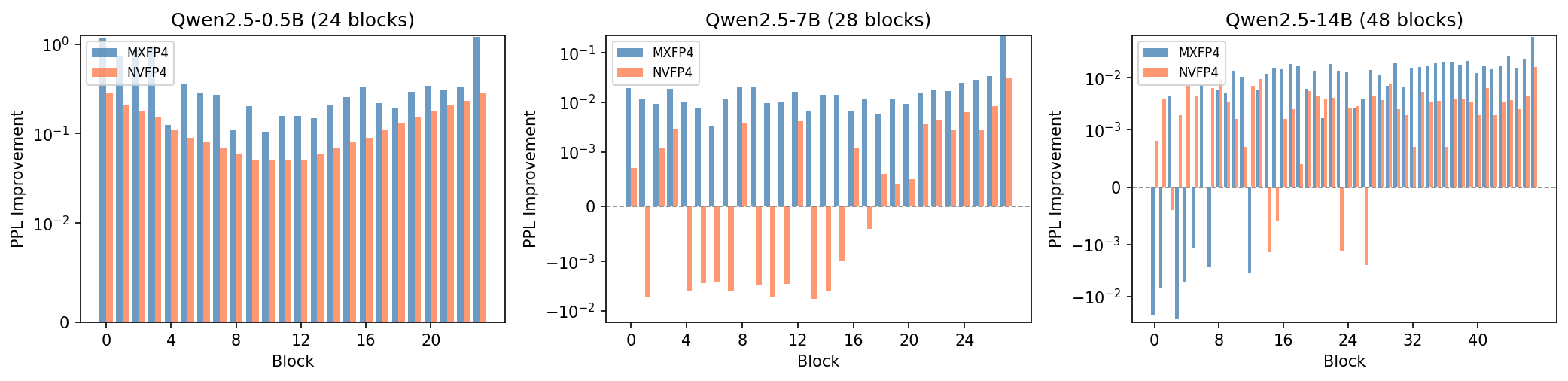}
\vspace{-5.5mm}
\caption{Block-wise sensitivity analysis for down projection across three model scales. Y-axis uses symlog scale. Positive values indicate PPL improvement when keeping that block in FP16.}
\label{fig:block_down_proj}
\vspace{-3mm}
\end{figure}

\noindent {\bf Block Sensitivity.} We analyze the FP4 sensitivity across transformer depth using block-wise isolation and observe that the sensitivity is structured rather than uniformly concentrated in the final blocks. While later blocks are often sensitive, particularly for MLP projections, early-block sensitivity can also be substantial, most clearly in the 0.5B model and under MXFP4. At larger scale, sensitivity is more concentrated towards later blocks, although early blocks can still exhibit non-trivial effects; notably, for 14B under MXFP4, isolating early blocks can even yield negative improvement. Figure~\ref{fig:block_down_proj} shows the block-wise perplexity when quantizing down projection across all three model scales.

\keytakeaway{\textbf{Key Takeaway \#2:}} FP4 sensitivity does not always peak in the final blocks and can vary across depths depending on the model network configuration.

\noindent {\bf Activation Outlier Analysis.} To understand the component sensitivity ordering, we analyze activation statistics as reported in Table~\ref{tab:outliers}. Down projection exhibits extreme outlier behavior with Max/P99.9 ratios 10-100$\times$ larger than other components, consistent with its high FP4 sensitivity. However, up projection shows comparable sensitivity despite much lower outlier ratios, indicating that outliers alone do not fully explain the FP4 sensitivity.

\begin{table}[ht]
\begin{center}
\small
\begin{tabular}{lccc|ccc|ccc}
\toprule
& \multicolumn{3}{c|}{\textbf{Qwen2.5-0.5B}} & \multicolumn{3}{c|}{\textbf{Qwen2.5-7B}} & \multicolumn{3}{c}{\textbf{Qwen2.5-14B}} \\
\cmidrule(lr){2-10}
\textbf{Component} & P99.9 & Max & Ratio & P99.9 & Max & Ratio & P99.9 & Max & Ratio \\
\midrule
\texttt{down\_proj} & 2.8 & 253 & \textbf{159} & 4.3 & 310 & \textbf{80} & 4.2 & 300 & \textbf{334} \\
\texttt{up\_proj} & 11.2 & 58 & 5.5 & 9.9 & 49 & 5.5 & 5.1 & 42 & 10 \\
\texttt{gate\_proj} & 11.2 & 58 & 5.5 & 9.9 & 49 & 5.5 & 5.1 & 46 & 11 \\
\texttt{v\_proj} & 13.0 & 36 & 2.9 & 7.0 & 30 & 4.4 & 4.8 & 23 & 4.8 \\
\texttt{o\_proj} & 2.4 & 6 & 2.6 & 2.4 & 9 & 3.4 & 1.8 & 6 & 3.6 \\
\texttt{k\_proj} & 13.0 & 36 & 2.9 & 7.0 & 29 & 4.3 & 4.8 & 24 & 4.9 \\
\texttt{q\_proj} & 13.0 & 36 & 2.9 & 7.0 & 29 & 4.4 & 4.8 & 22 & 4.7 \\
\bottomrule
\end{tabular}
\end{center}
\vspace{-1mm}
\caption{Activation statistics by component across all model scales. All values are averaged across layers. The Max/P99.9 ratio is computed as the mean of per-layer ratios (not the ratio of averaged Max to averaged P99.9), which better captures outlier severity across depth. Down projection, as highlighted, consistently shows substantially worse ratios than other components.}
\label{tab:outliers}
\end{table}

\keytakeaway{\textbf{Key Takeaway \#3:}} Extreme activation outliers in down projection, which receives post-activation values (e.g., after SiLU/SwiGLU gating~\citep{glu_mlp}), are consistent with its high sensitivity, but up projection exhibits comparable sensitivity despite much lower outlier ratios, suggesting that extreme outliers alone do not fully account for FP4 sensitivity.

\section{Conclusion and Discussion} \vspace{-2.5mm}
We presented a controlled component-wise and block-wise analysis of FP4 quantization in transformer models across multiple formats and scales. Our results show that FP4 sensitivity is dominated by MLP projections with a stable tiering across formats and model sizes, while model scale primarily affects sensitivity magnitude. At the block level, sensitivity is structured across depth and not exclusively concentrated in the final blocks, with early-block effects emerging in some configurations. Together, these findings provide a diagnostic view of FP4 inference behavior and motivate component- and depth-aware approaches to low-precision deployment. Looking ahead, this analysis can be extended to additional model families and larger scales, as well as to settings that employ native FP4 computation kernels. Future work should also evaluate FP4 sensitivity on diverse downstream tasks beyond perplexity on WikiText, such as reasoning, coding, and instruction following benchmarks, to better understand how component level quantization effects translate to task specific performance degradation.

\newpage

\bibliographystyle{iclr2026_conference}
{\footnotesize
\setlength{\bibsep}{0pt plus 0.3ex}
\bibliography{iclr2026_conference}
}

\appendix

\section{Related Work}

Prior work on efficient large language model deployment has primarily focused on higher-precision regimes, including FP16 and FP8. Foundational model families such as LLaMA-2 and LLaMA-3 are trained and deployed in FP16, establishing strong baselines for accuracy and scaling behavior \cite{touvron2023llama,grattafiori2024llama}. More recent work explores FP8 training and inference, showing that carefully designed scaling, normalization, and GEMM kernels can enable stable low-precision execution at scale \cite{peng2023fp8,hernandez2025towards}. These approaches largely treat quantization as a uniform transformation across model components and layers, without examining how sensitivity varies within the transformer architecture.

Recent studies have pushed further toward ultra-low-precision FP4 formats to maximize efficiency. Prior work proposes microscaling and mixed-precision techniques for FP4 inference and training, including NVFP4 and MXFP4-based approaches, residual channels, and hybrid precision strategies \cite{egiazarian2025bridging,tseng2025training,abecassis2025pretraining,zhao2024atom,meng2026arcquant,liu2025micromix}. More recently, SageAttention3 investigates microscaling FP4 specifically for attention inference and explores interactions with 8-bit training \cite{zhang2025sageattention3}, highlighting that different architectural components may exhibit distinct FP4 sensitivity. In contrast to these method-driven efforts, our work provides a diagnostic component-wise and block-wise analysis of FP4 sensitivity, characterizing how quantization effects distribute across components and depth rather than proposing a new quantization scheme.

\section{Component Sensitivity Tables}

\begin{table}[H]
\caption{Component sensitivity under NVFP4 and MXFP4 for Qwen2.5-0.5B. PPL improvement measures sensitivity (higher = keeping that component in FP16 helps more).}
\label{tab:component}
\begin{center}
\small
\begin{tabular}{lcccc}
\toprule
& \multicolumn{2}{c}{\textbf{NVFP4}} & \multicolumn{2}{c}{\textbf{MXFP4}} \\
\cmidrule(lr){2-3} \cmidrule(lr){4-5}
\textbf{Component} & PPL & Improvement & PPL & Improvement \\
\midrule
Baseline & 21.63 & --- & 36.71 & --- \\
\midrule
\texttt{up\_proj} & 20.40 & +1.23 & 28.46 & +8.25 \\
\texttt{down\_proj} & 20.47 & +1.16 & 28.79 & +7.92 \\
\texttt{gate\_proj} & 20.73 & +0.90 & 31.24 & +5.47 \\
\texttt{v\_proj} & 20.99 & +0.64 & 31.45 & +5.26 \\
\texttt{o\_proj} & 21.04 & +0.59 & 31.77 & +4.94 \\
\texttt{k\_proj} & 21.29 & +0.34 & 33.46 & +3.25 \\
\texttt{q\_proj} & 21.37 & +0.26 & 34.73 & +1.98 \\
\bottomrule
\end{tabular}
\end{center}
\end{table}

\begin{table}[H]
\caption{Component sensitivity under NVFP4 and MXFP4 for Qwen2.5-7B.}
\label{tab:component_7b}
\begin{center}
\small
\begin{tabular}{lcccc}
\toprule
& \multicolumn{2}{c}{\textbf{NVFP4}} & \multicolumn{2}{c}{\textbf{MXFP4}} \\
\cmidrule(lr){2-3} \cmidrule(lr){4-5}
\textbf{Component} & PPL & Improvement & PPL & Improvement \\
\midrule
Baseline & 6.47 & --- & 7.31 & --- \\
\midrule
\texttt{down\_proj} & 6.36 & +0.11 & 6.81 & +0.50 \\
\texttt{up\_proj} & 6.37 & +0.10 & 6.99 & +0.31 \\
\texttt{gate\_proj} & 6.39 & +0.08 & 7.08 & +0.23 \\
\texttt{o\_proj} & 6.43 & +0.04 & 7.18 & +0.13 \\
\texttt{v\_proj} & 6.44 & +0.03 & 7.15 & +0.15 \\
\texttt{k\_proj} & 6.45 & +0.02 & 7.21 & +0.10 \\
\texttt{q\_proj} & 6.45 & +0.02 & 7.23 & +0.07 \\
\bottomrule
\end{tabular}
\end{center}
\end{table}

\vspace{-1em}
\begin{table}[H]
\caption{Component sensitivity under NVFP4 and MXFP4 for Qwen2.5-14B.}
\label{tab:component_14b}
\small
\begin{center}
\begin{tabular}{lcccc}
\toprule
& \multicolumn{2}{c}{\textbf{NVFP4}} & \multicolumn{2}{c}{\textbf{MXFP4}} \\
\cmidrule(lr){2-3} \cmidrule(lr){4-5}
\textbf{Component} & PPL & Improvement & PPL & Improvement \\
\midrule
Baseline & 5.26 & --- & 6.02 & --- \\
\midrule
\texttt{down\_proj} & 5.10 & +0.16 & 5.63 & +0.39 \\
\texttt{up\_proj} & 5.12 & +0.14 & 5.77 & +0.24 \\
\texttt{gate\_proj} & 5.14 & +0.12 & 5.82 & +0.20 \\
\texttt{o\_proj} & 5.21 & +0.05 & 5.89 & +0.13 \\
\texttt{v\_proj} & 5.23 & +0.03 & 5.95 & +0.07 \\
\texttt{k\_proj} & 5.24 & +0.02 & 5.92 & +0.10 \\
\texttt{q\_proj} & 5.24 & +0.02 & 5.94 & +0.08 \\
\bottomrule
\end{tabular}
\end{center}
\end{table}

\section{Block Sensitivity Analysis (Qwen2.5-0.5B)}

This section presents detailed block sensitivity analysis for the Qwen2.5-0.5B model (24 blocks). Each component shows two figures: (1) raw perplexity values, and (2) percentage change from baseline. Negative percentages indicate improvement (lower PPL), with 0\% representing the baseline.

\begin{figure}[H]
\begin{center}
\includegraphics[width=\linewidth]{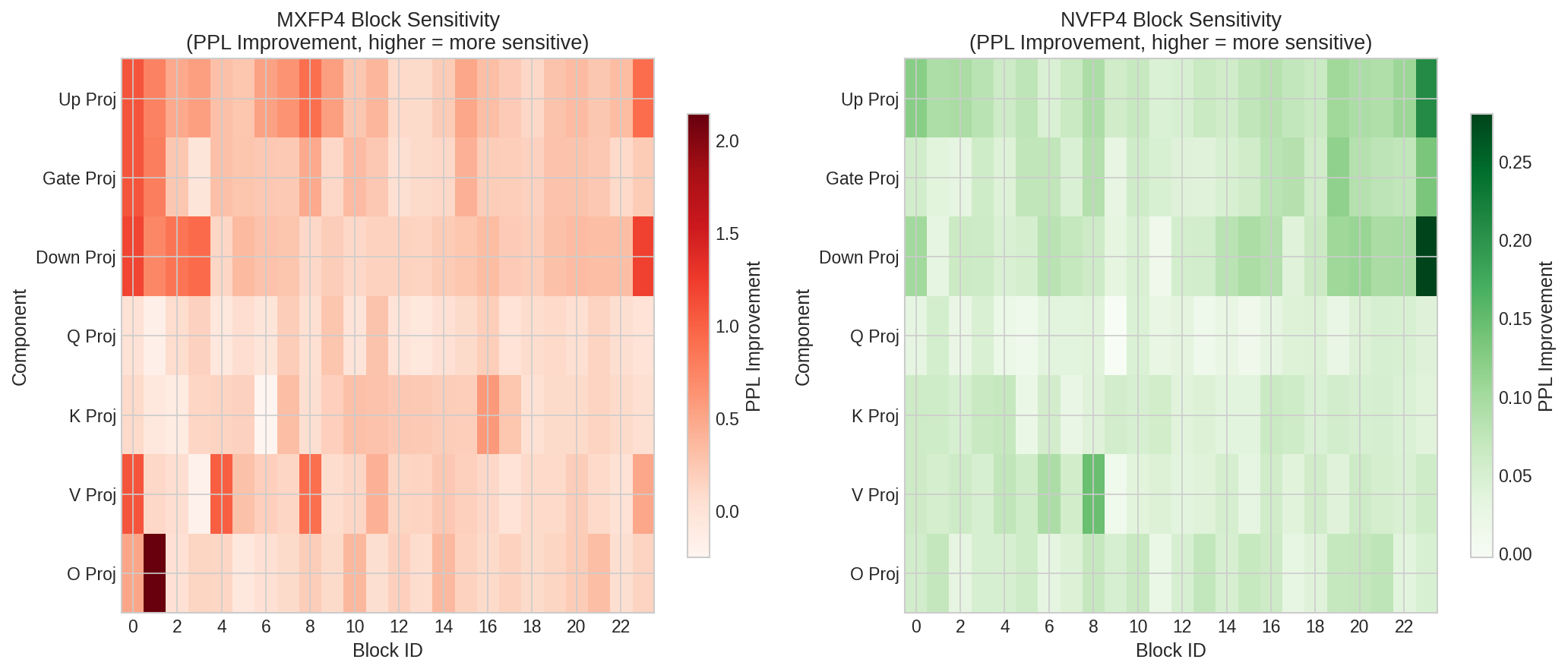}
\end{center}
\caption{Block sensitivity heatmaps showing PPL improvement when each block's component is kept in FP16. Left: MXFP4 (scale 0--2.0). Right: NVFP4 (scale 0--0.28). Note the 7$\times$ scale difference and different spatial patterns.}
\label{fig:heatmap}
\end{figure}

\subsection{MLP Components}

\begin{figure}[H]
\begin{center}
\includegraphics[width=0.85\linewidth]{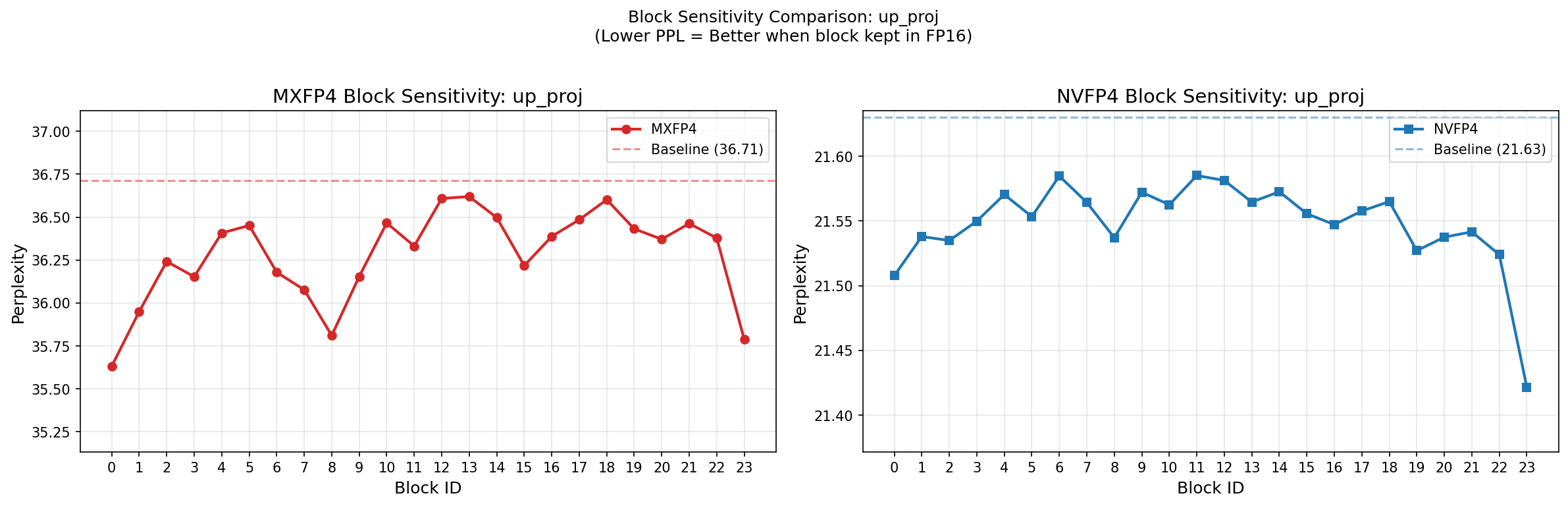}
\end{center}
\vspace{-2.5mm}
\caption{Block sensitivity for \texttt{up\_proj}. MXFP4 shows strong early-block sensitivity (blocks 0, 8, 23), while NVFP4 peaks at block 23.}
\label{fig:block_up_proj}
\end{figure}

\begin{figure}[H]
\begin{center}
\includegraphics[width=0.85\linewidth]{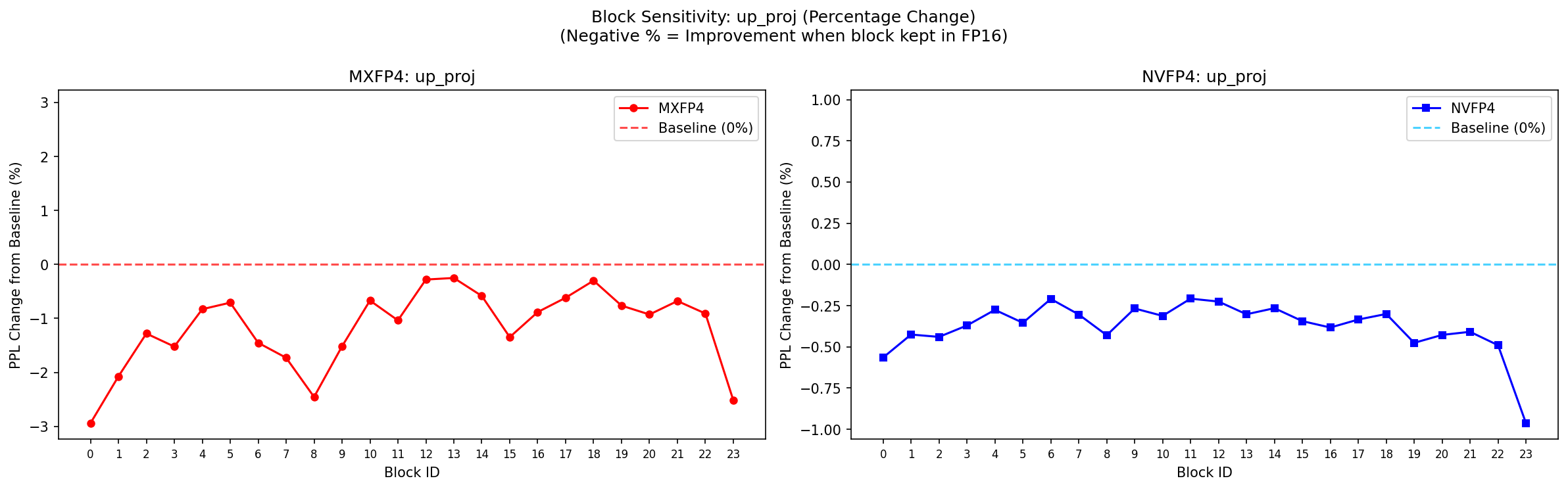}
\end{center}
\vspace{-2.5mm}
\caption{Percentage change from baseline for \texttt{up\_proj}. Block 23 shows $-3.3\%$ (MXFP4) and $-1.3\%$ (NVFP4) improvement.}
\label{fig:block_up_proj_pct}
\end{figure}

\begin{figure}[H]
\begin{center}
\includegraphics[width=0.85\linewidth]{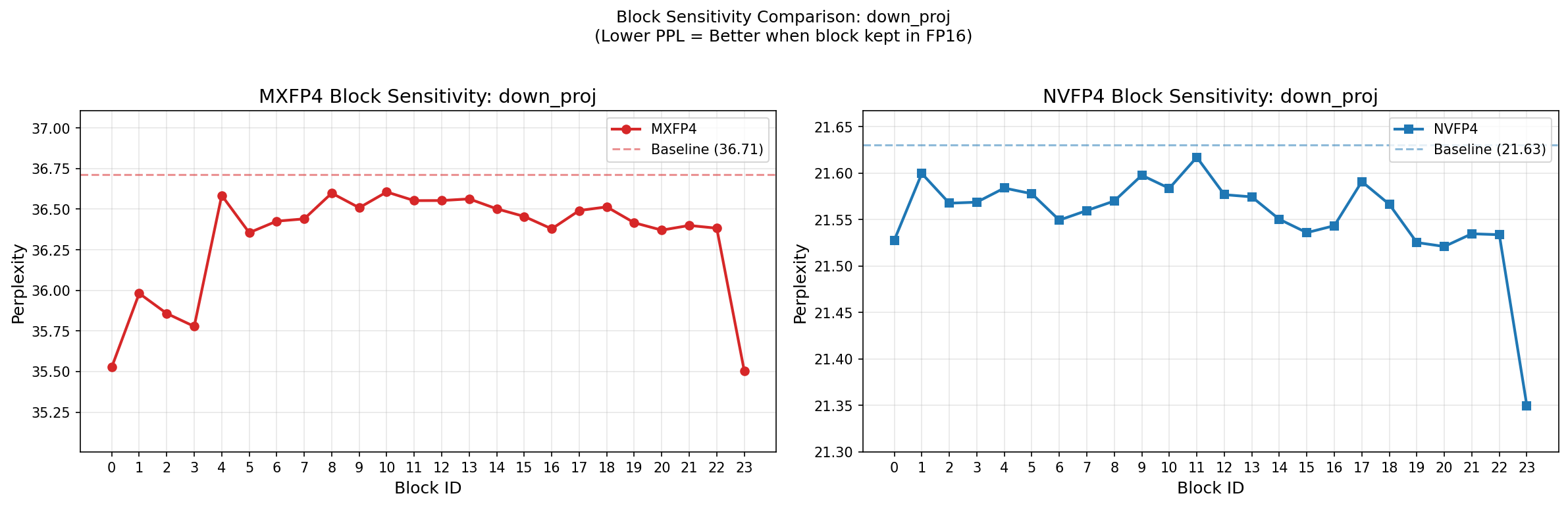}
\end{center}
\vspace{-2.5mm}
\caption{Block sensitivity for \texttt{down\_proj} (0.5B only). Both formats show block 23 as highly sensitive, but MXFP4 also shows strong sensitivity in blocks 0--3.}
\label{fig:block_down_proj_05b}
\end{figure}

\begin{figure}[H]
\begin{center}
\includegraphics[width=0.85\linewidth]{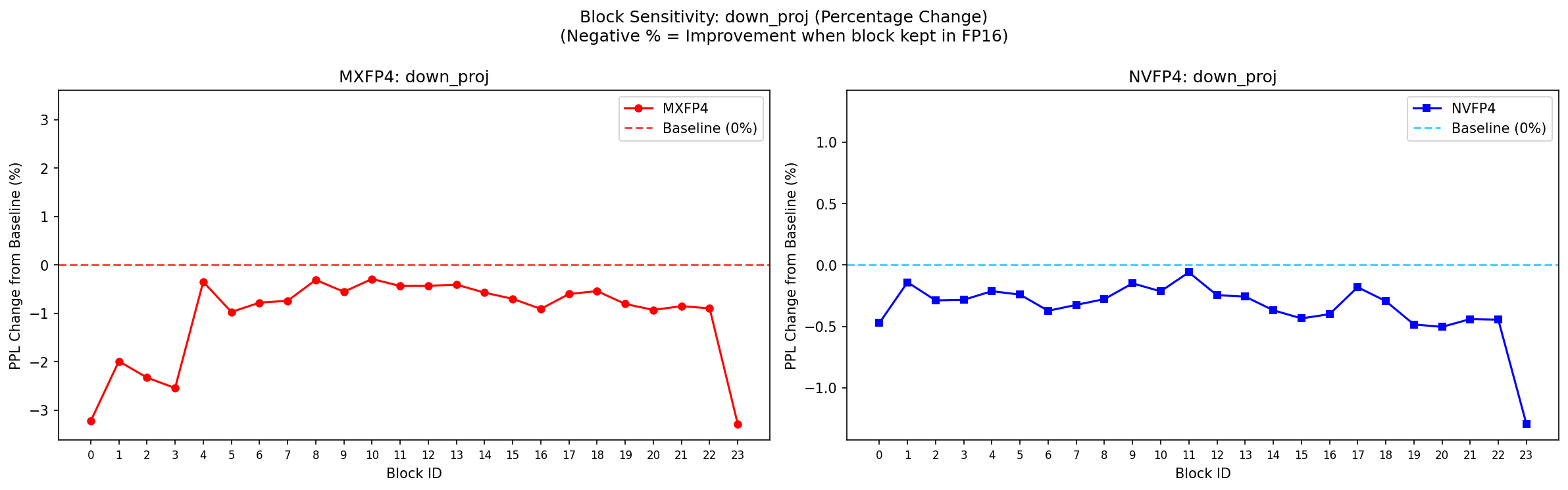}
\end{center}
\vspace{-2.5mm}
\caption{Percentage change from baseline for \texttt{down\_proj}. Block 23 shows $-3.3\%$ (MXFP4) and $-1.3\%$ (NVFP4) improvement.}
\label{fig:block_down_proj_pct}
\end{figure}

\begin{figure}[H]
\begin{center}
\includegraphics[width=0.85\linewidth]{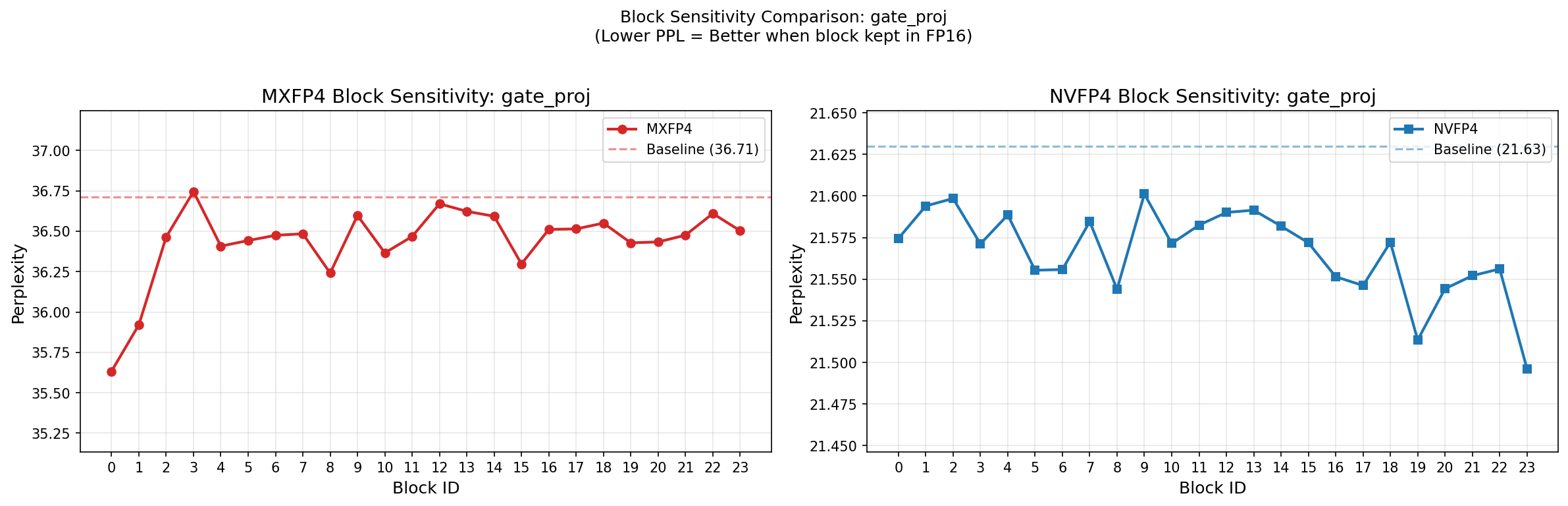}
\end{center}
\vspace{-2.5mm}
\caption{Block sensitivity for \texttt{gate\_proj}. MXFP4 exhibits pronounced early-block dominance (blocks 0--2), while NVFP4 shows late-block sensitivity peaking at block 23.}
\label{fig:block_gate_proj}
\end{figure}

\begin{figure}[H]
\begin{center}
\includegraphics[width=0.85\linewidth]{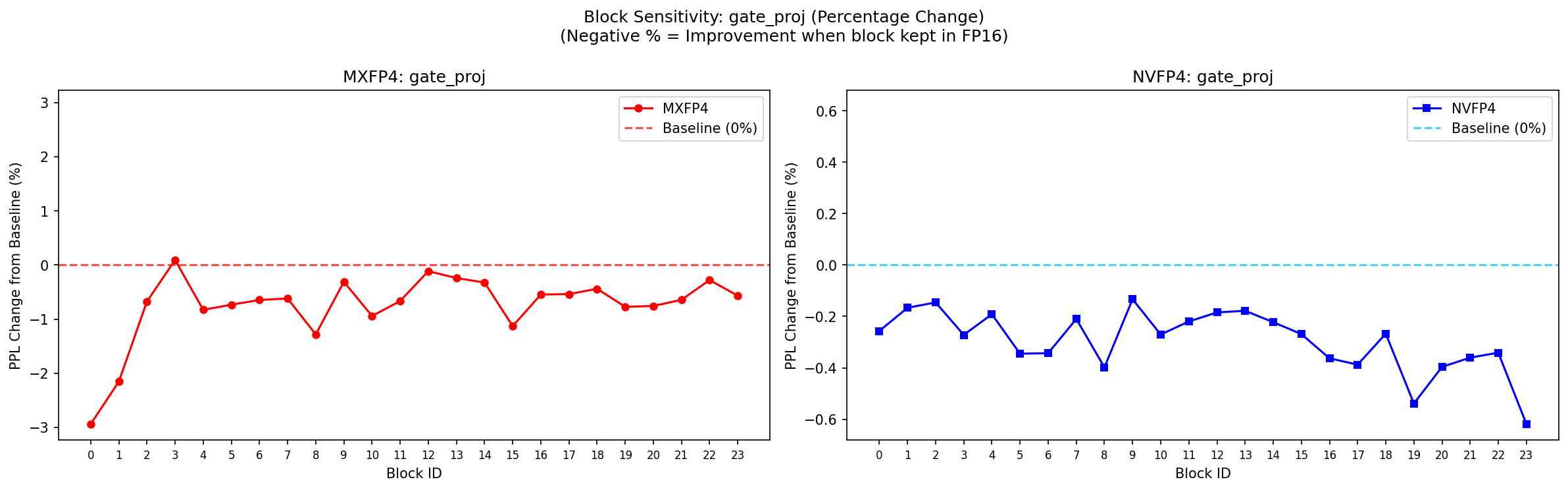}
\end{center}
\vspace{-2.5mm}
\caption{Percentage change from baseline for \texttt{gate\_proj}.}
\label{fig:block_gate_proj_pct}
\end{figure}

\FloatBarrier
\subsection{Attention Components}

\begin{figure}[H]
\begin{center}
\includegraphics[width=0.85\linewidth]{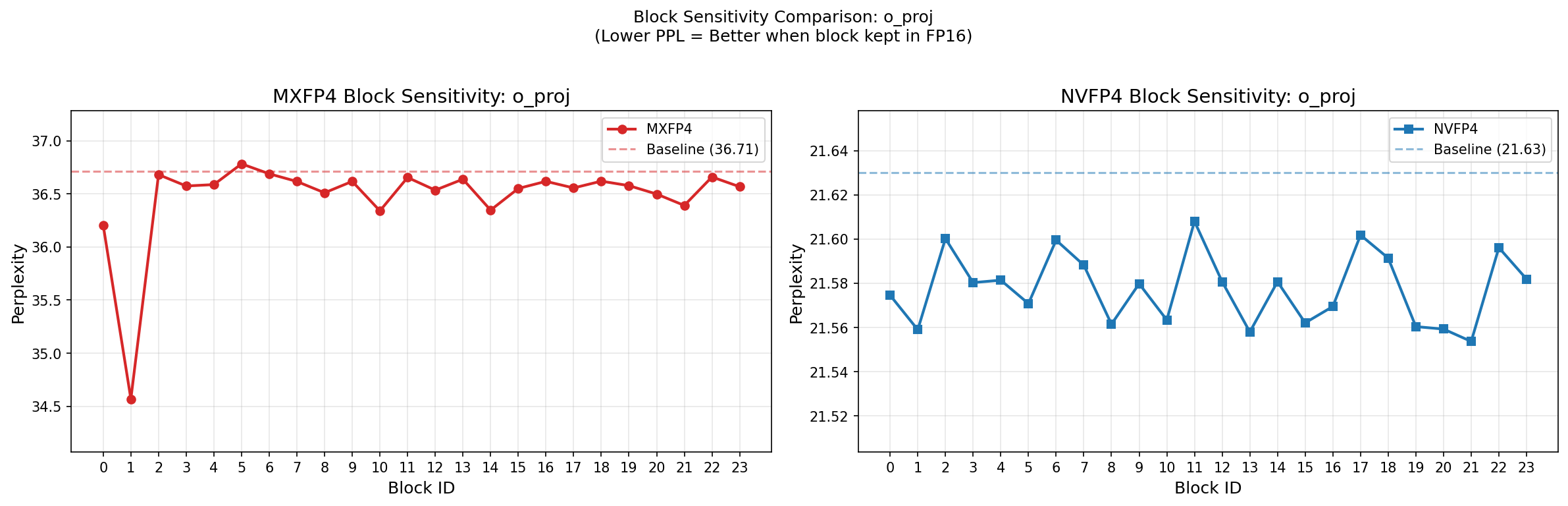}
\end{center}
\vspace{-2.5mm}
\caption{Block sensitivity for \texttt{o\_proj}. MXFP4 shows extreme sensitivity in block 1 (+2.14 PPL improvement), a unique pattern not seen in other components.}
\label{fig:block_o_proj}
\end{figure}

\begin{figure}[H]
\begin{center}
\includegraphics[width=0.85\linewidth]{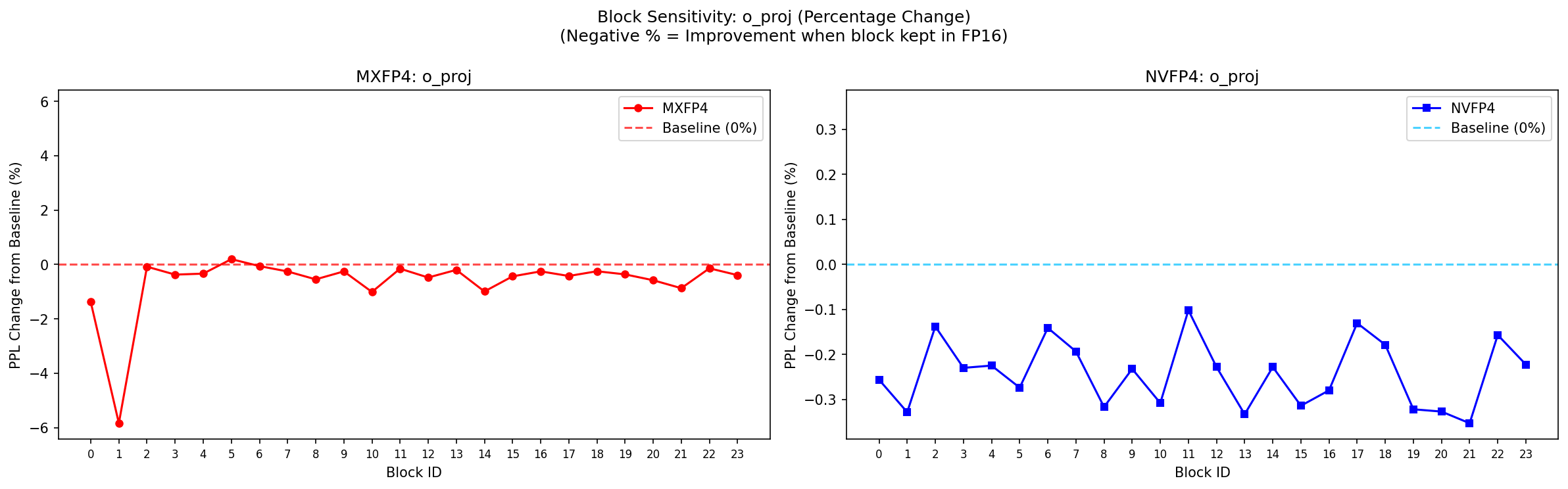}
\end{center}
\vspace{-2.5mm}
\caption{Percentage change from baseline for \texttt{o\_proj}. Block 1 shows $-5.8\%$ improvement for MXFP4.}
\label{fig:block_o_proj_pct}
\end{figure}

\begin{figure}[H]
\begin{center}
\includegraphics[width=0.85\linewidth]{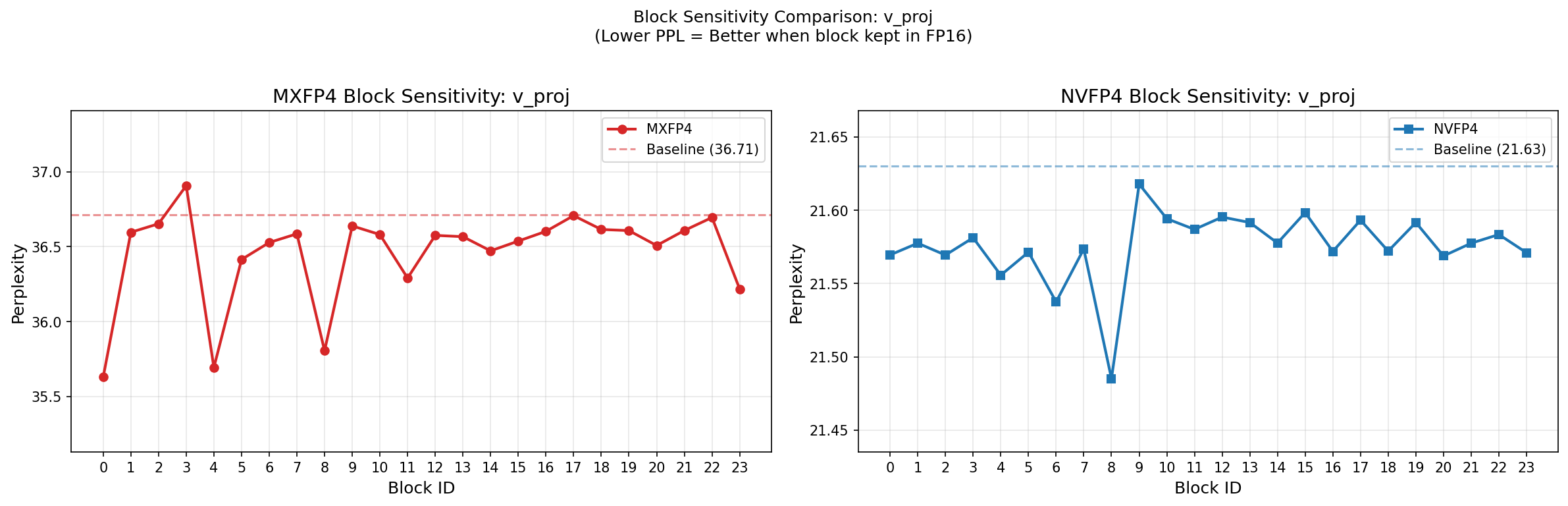}
\end{center}
\vspace{-2.5mm}
\caption{Block sensitivity for \texttt{v\_proj}. MXFP4 shows early-block dominance, while NVFP4 exhibits relatively flat sensitivity with a slight peak at block 8.}
\label{fig:block_v_proj}
\end{figure}

\begin{figure}[H]
\begin{center}
\includegraphics[width=0.85\linewidth]{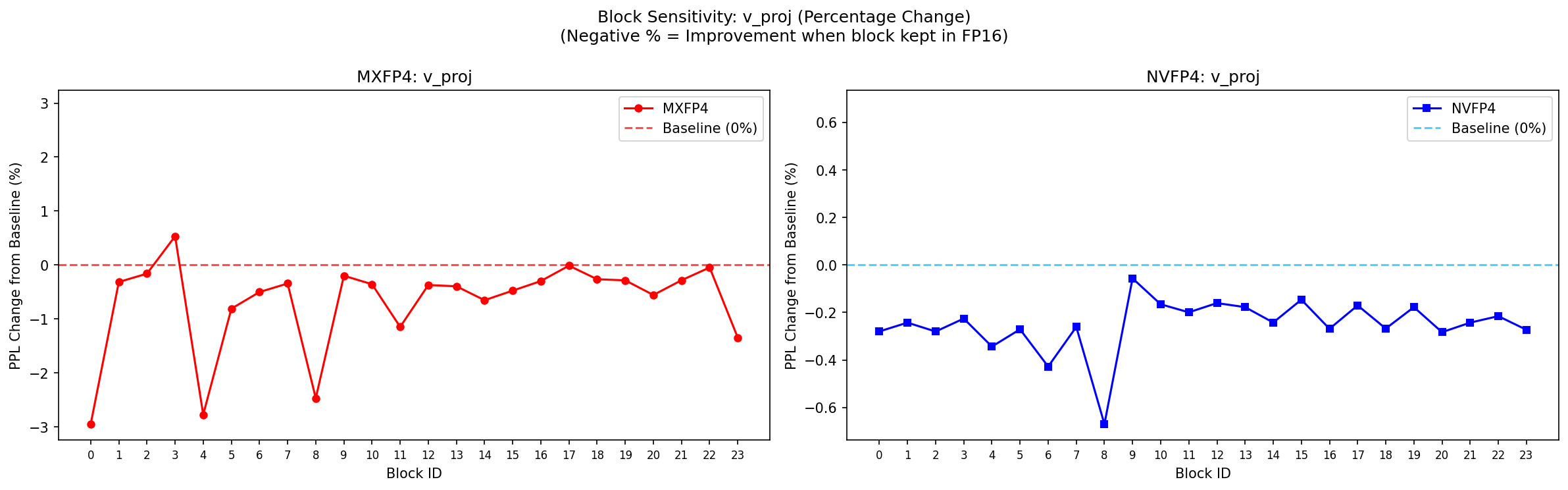}
\end{center}
\vspace{-2.5mm}
\caption{Percentage change from baseline for \texttt{v\_proj}.}
\label{fig:block_v_proj_pct}
\end{figure}

\begin{figure}[H]
\begin{center}
\includegraphics[width=0.85\linewidth]{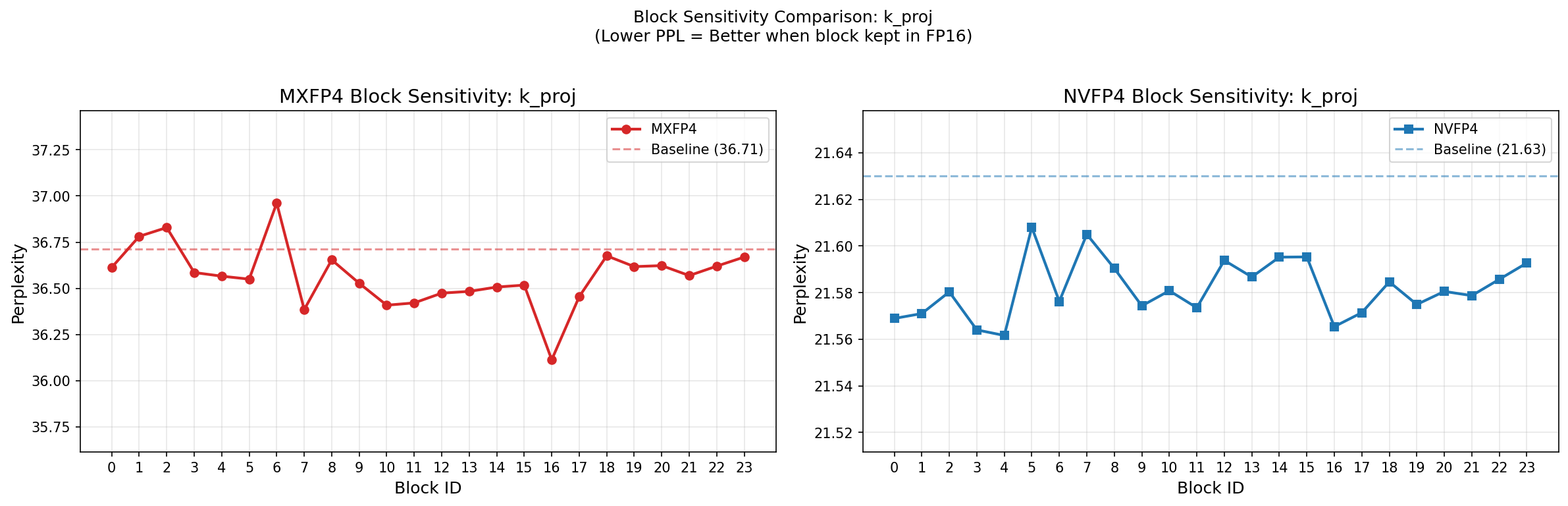}
\end{center}
\vspace{-2.5mm}
\caption{Block sensitivity for \texttt{k\_proj}. Both formats show relatively low and uniform sensitivity, consistent with \texttt{k\_proj} being the second-least sensitive component.}
\label{fig:block_k_proj}
\end{figure}

\begin{figure}[H]
\begin{center}
\includegraphics[width=0.85\linewidth]{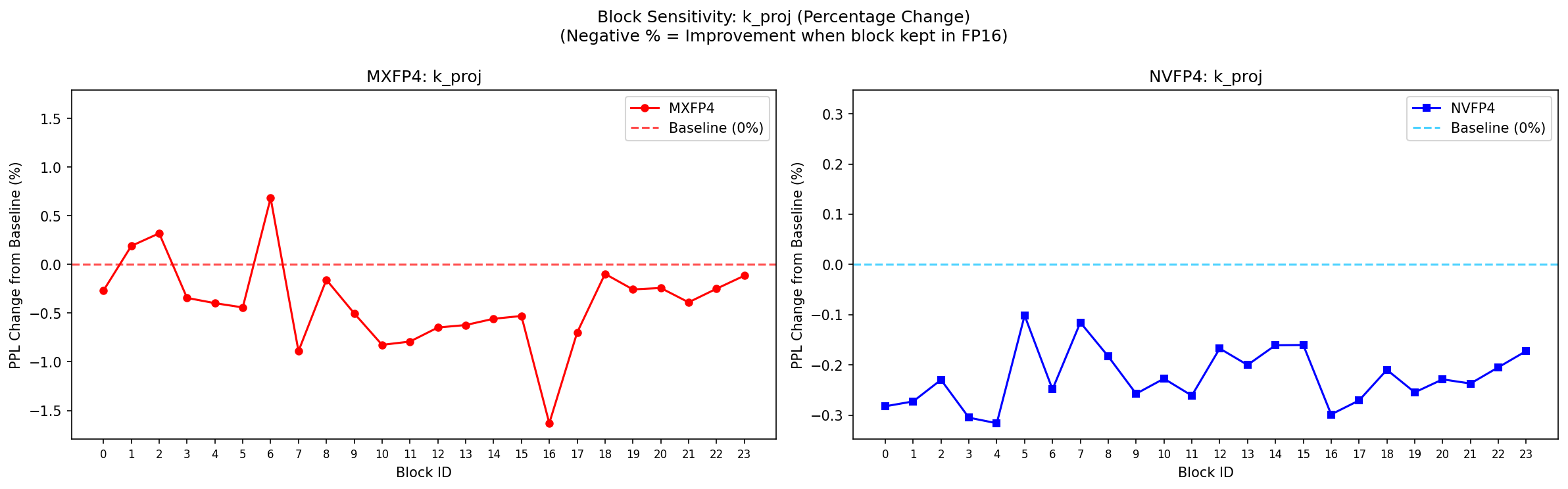}
\end{center}
\vspace{-2.5mm}
\caption{Percentage change from baseline for \texttt{k\_proj}.}
\label{fig:block_k_proj_pct}
\end{figure}

\begin{figure}[H]
\begin{center}
\includegraphics[width=0.85\linewidth]{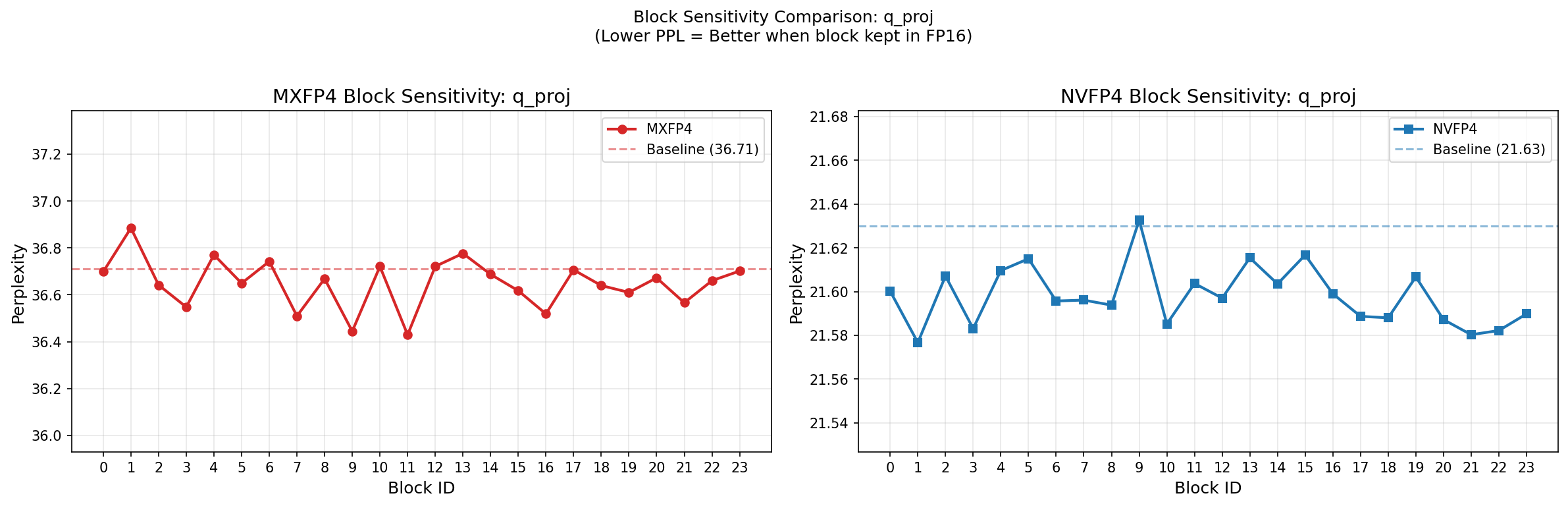}
\end{center}
\vspace{-2.5mm}
\caption{Block sensitivity for \texttt{q\_proj}. The least sensitive component overall, showing minimal variation across blocks for both formats.}
\label{fig:block_q_proj}
\end{figure}

\begin{figure}[H]
\begin{center}
\includegraphics[width=0.85\linewidth]{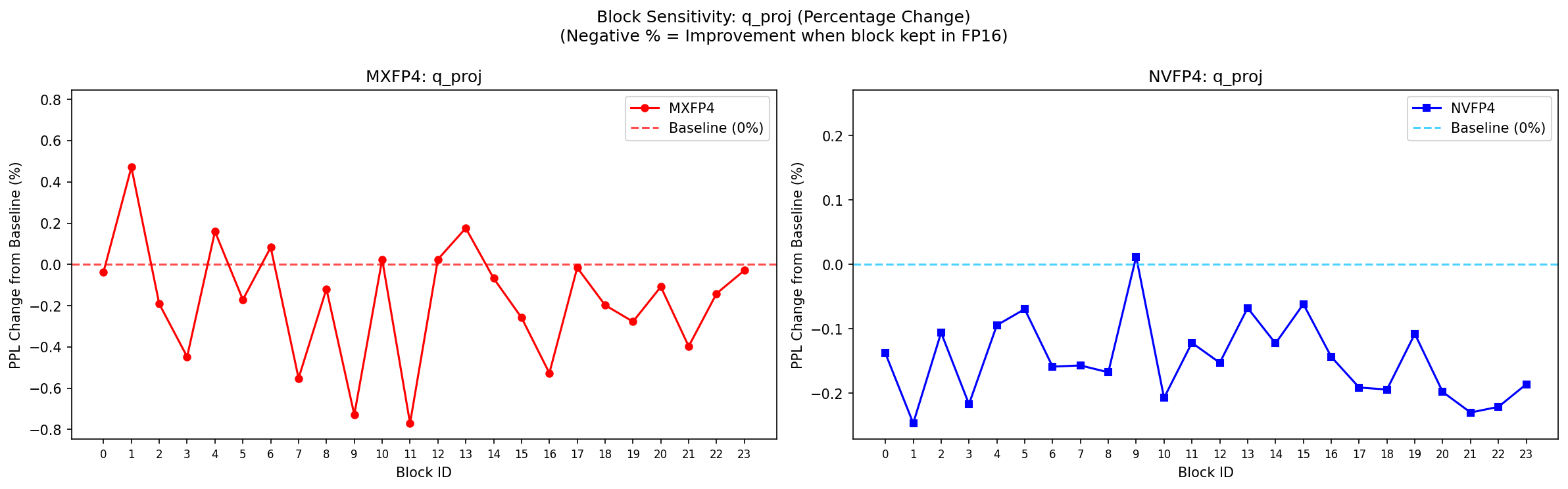}
\end{center}
\vspace{-2.5mm}
\caption{Percentage change from baseline for \texttt{q\_proj}.}
\label{fig:block_q_proj_pct}
\end{figure}

\section{Block Sensitivity Analysis (Qwen2.5-7B)}

This section presents detailed block sensitivity analysis for the Qwen2.5-7B model (28 blocks). Each component shows two figures: (1) raw perplexity values, and (2) percentage change from baseline. Negative percentages indicate improvement (lower PPL), with 0\% representing the baseline.

\begin{figure}[H]
\begin{center}
\includegraphics[width=\linewidth]{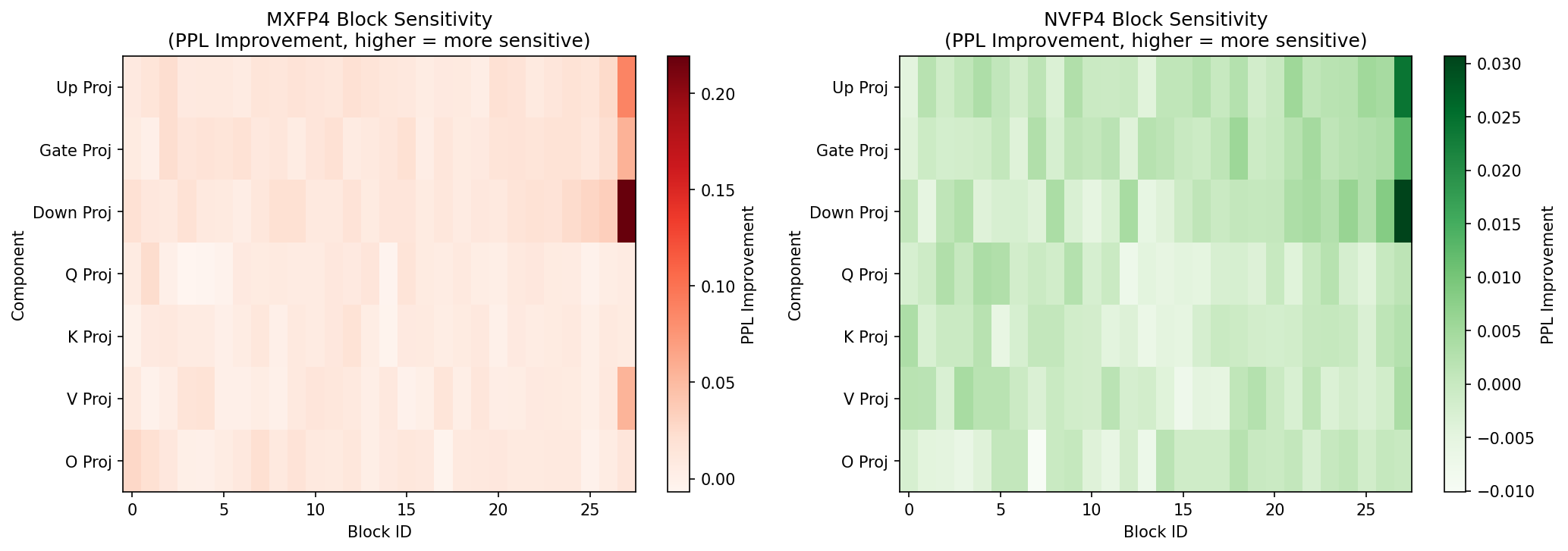}
\end{center}
\caption{Block sensitivity heatmaps for Qwen2.5-7B showing PPL improvement when each block's component is kept in FP16. Left: MXFP4. Right: NVFP4. Block 27 shows strong sensitivity for MLP components in both formats.}
\label{fig:heatmap_7b}
\end{figure}

\subsection{MLP Components}

\begin{figure}[H]
\begin{center}
\includegraphics[width=0.85\linewidth]{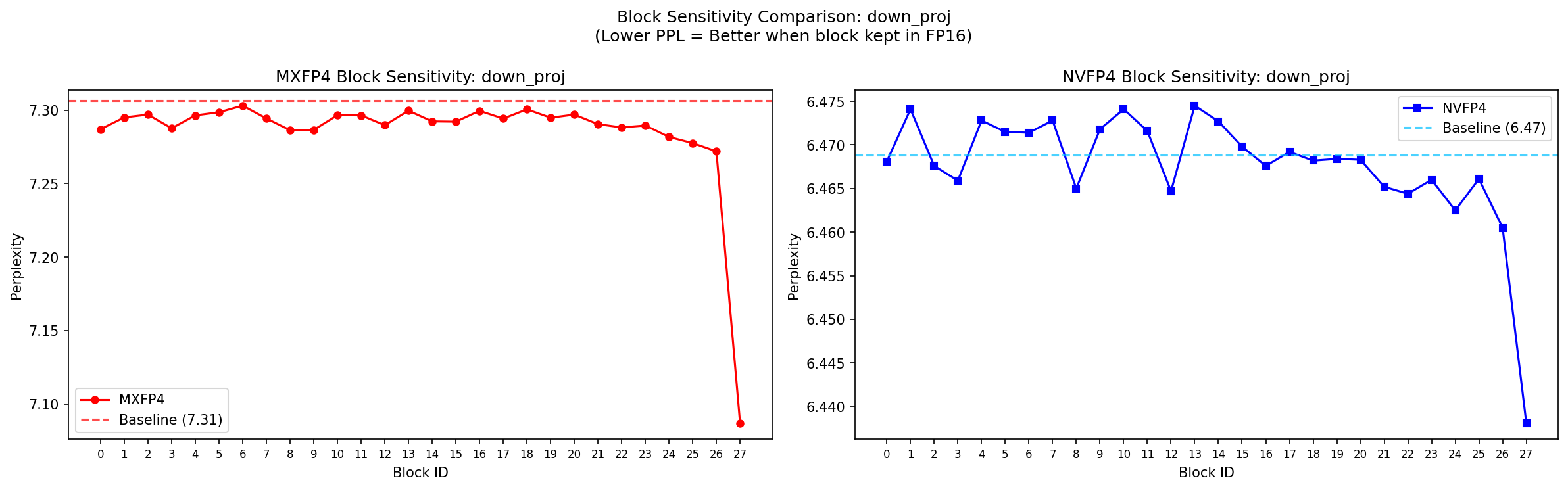}
\end{center}
\vspace{-2.5mm}
\caption{Block sensitivity for \texttt{down\_proj} (7B). Both formats show block 27 as highly sensitive.}
\label{fig:block_down_proj_7b}
\end{figure}

\begin{figure}[H]
\begin{center}
\includegraphics[width=0.85\linewidth]{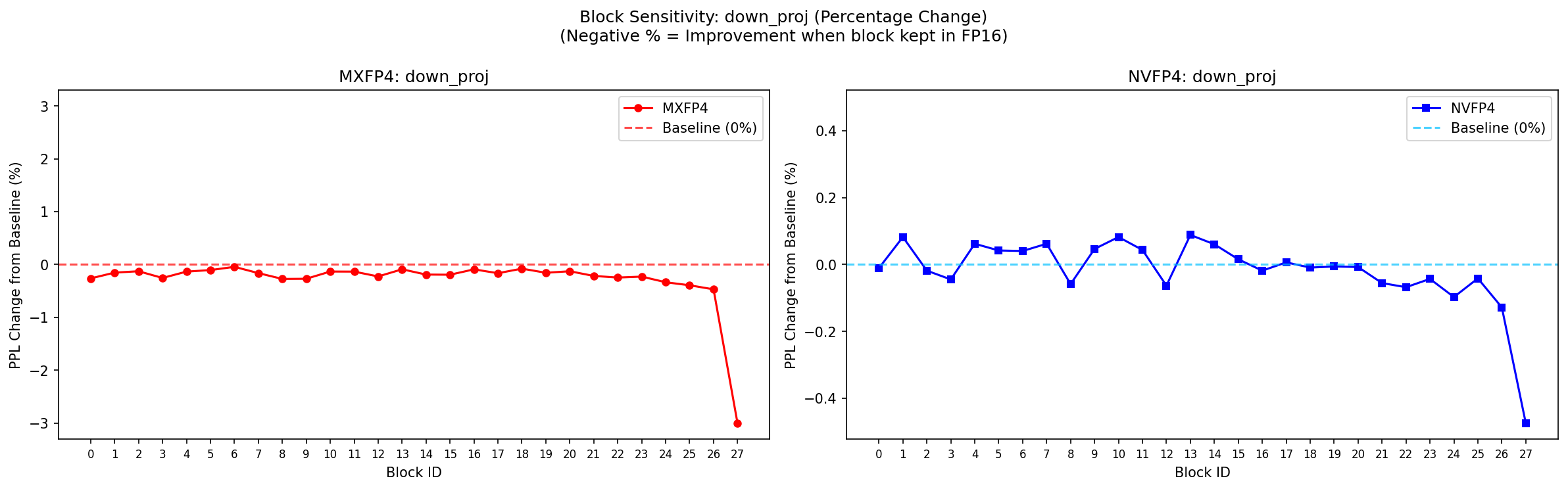}
\end{center}
\vspace{-2.5mm}
\caption{Percentage change from baseline for \texttt{down\_proj} (7B). Block 27 shows $-3.0\%$ (MXFP4) and $-0.47\%$ (NVFP4) improvement.}
\label{fig:block_down_proj_7b_pct}
\end{figure}

\begin{figure}[H]
\begin{center}
\includegraphics[width=0.85\linewidth]{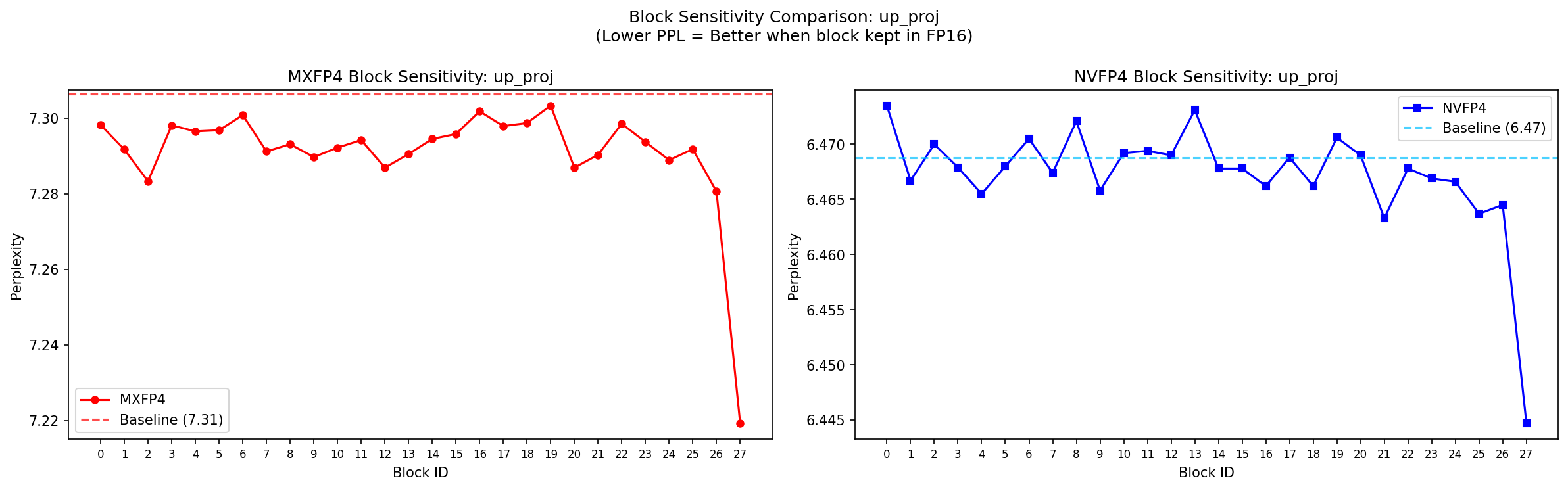}
\end{center}
\vspace{-2.5mm}
\caption{Block sensitivity for \texttt{up\_proj} (7B). Block 27 is most sensitive for both formats.}
\label{fig:block_up_proj_7b}
\end{figure}

\begin{figure}[H]
\begin{center}
\includegraphics[width=0.85\linewidth]{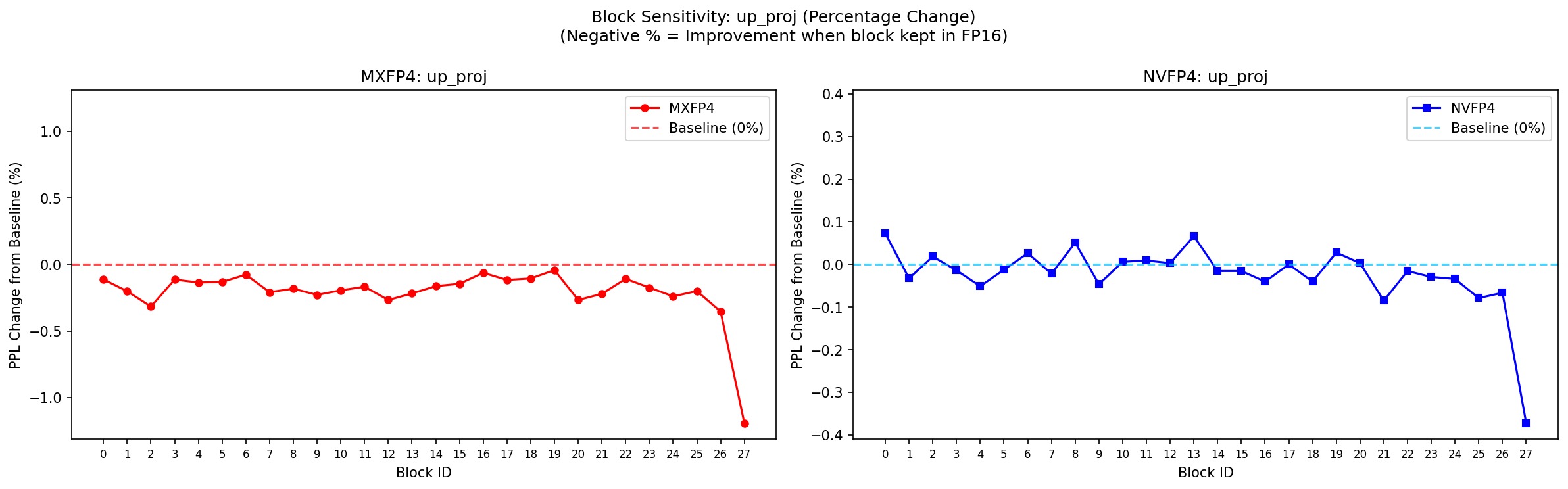}
\end{center}
\vspace{-2.5mm}
\caption{Percentage change from baseline for \texttt{up\_proj} (7B). Block 27 shows $-1.2\%$ (MXFP4) and $-0.37\%$ (NVFP4) improvement.}
\label{fig:block_up_proj_7b_pct}
\end{figure}

\begin{figure}[H]
\begin{center}
\includegraphics[width=0.85\linewidth]{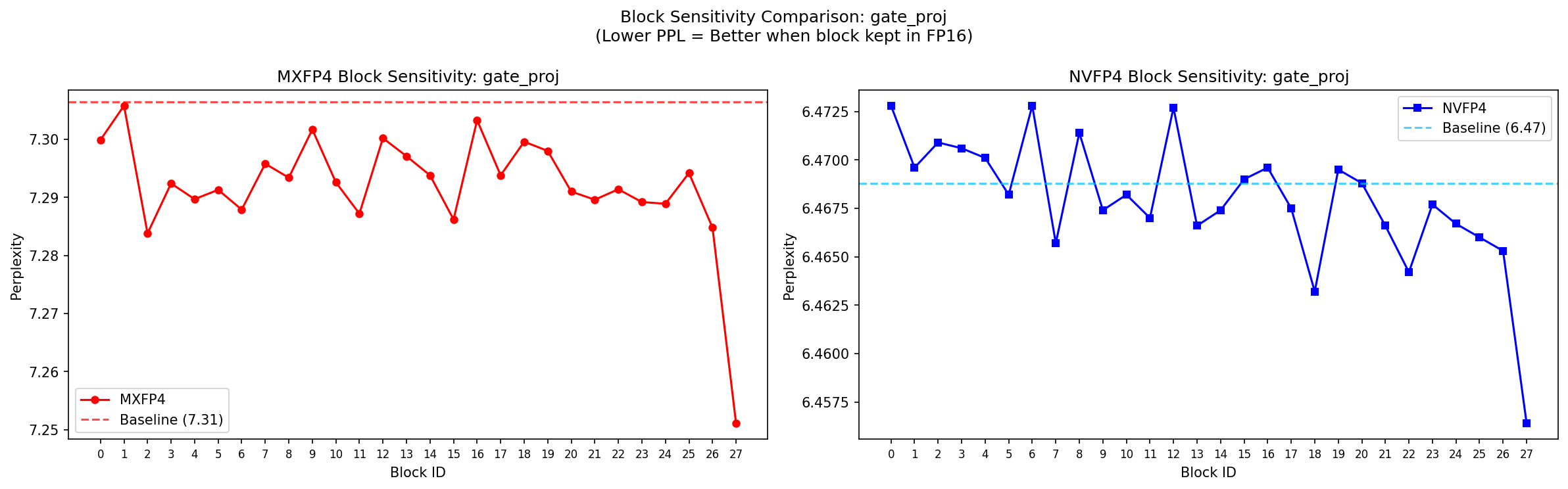}
\end{center}
\vspace{-2.5mm}
\caption{Block sensitivity for \texttt{gate\_proj} (7B). Block 27 shows highest sensitivity.}
\label{fig:block_gate_proj_7b}
\end{figure}

\begin{figure}[H]
\begin{center}
\includegraphics[width=0.85\linewidth]{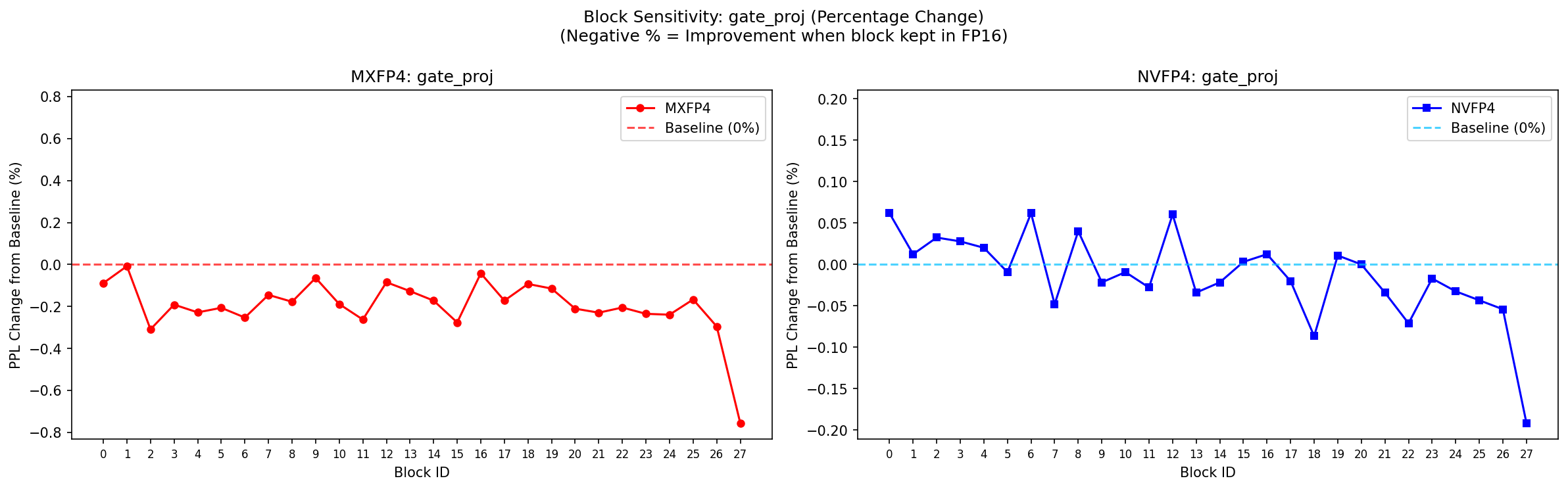}
\end{center}
\vspace{-2.5mm}
\caption{Percentage change from baseline for \texttt{gate\_proj} (7B). Block 27 shows $-0.76\%$ (MXFP4) and $-0.19\%$ (NVFP4) improvement.}
\label{fig:block_gate_proj_7b_pct}
\end{figure}

\FloatBarrier
\subsection{Attention Components}

\begin{figure}[H]
\begin{center}
\includegraphics[width=0.85\linewidth]{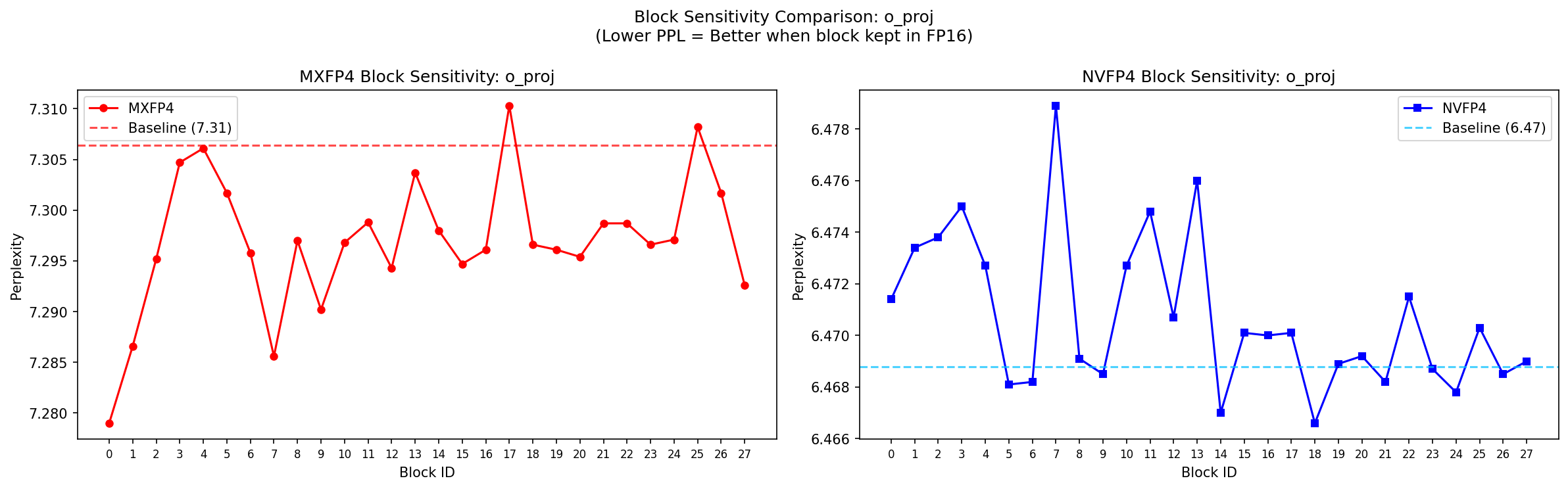}
\end{center}
\vspace{-2.5mm}
\caption{Block sensitivity for \texttt{o\_proj} (7B). MXFP4 shows early-block sensitivity, while NVFP4 shows relatively flat sensitivity.}
\label{fig:block_o_proj_7b}
\end{figure}

\begin{figure}[H]
\begin{center}
\includegraphics[width=0.85\linewidth]{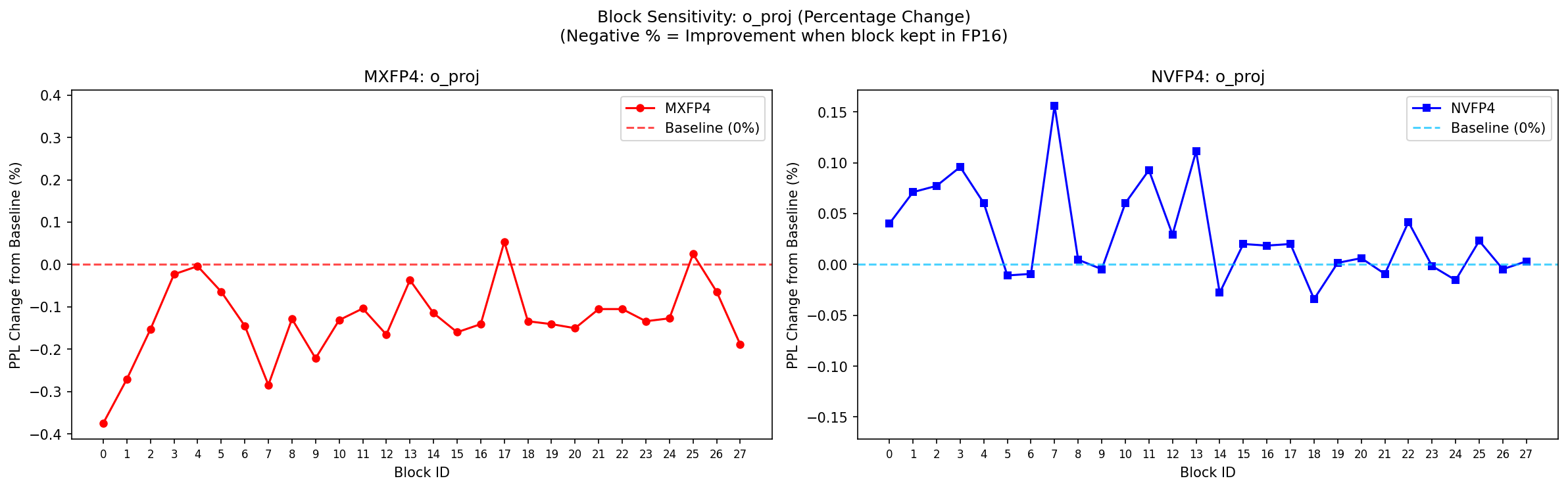}
\end{center}
\vspace{-2.5mm}
\caption{Percentage change from baseline for \texttt{o\_proj} (7B). MXFP4 shows $\sim$0.3\% variation, while NVFP4 shows $<$0.05\%.}
\label{fig:block_o_proj_7b_pct}
\end{figure}

\begin{figure}[H]
\begin{center}
\includegraphics[width=0.85\linewidth]{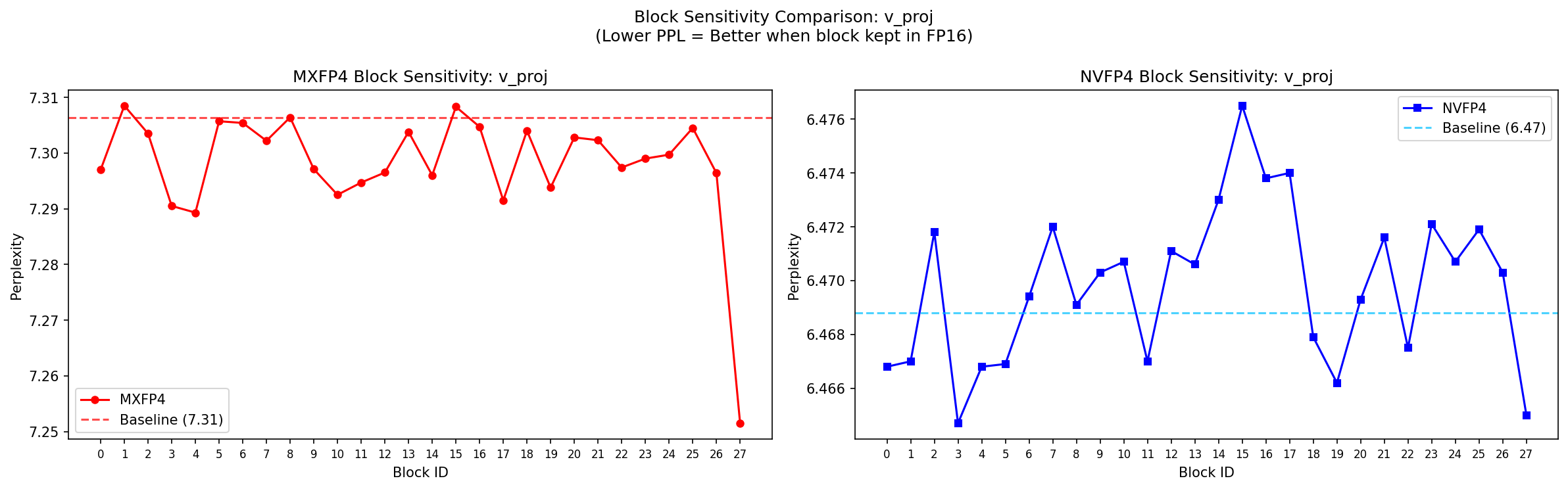}
\end{center}
\vspace{-2.5mm}
\caption{Block sensitivity for \texttt{v\_proj} (7B). MXFP4 shows block 27 as most sensitive, while NVFP4 exhibits relatively flat sensitivity.}
\label{fig:block_v_proj_7b}
\end{figure}

\begin{figure}[H]
\begin{center}
\includegraphics[width=0.85\linewidth]{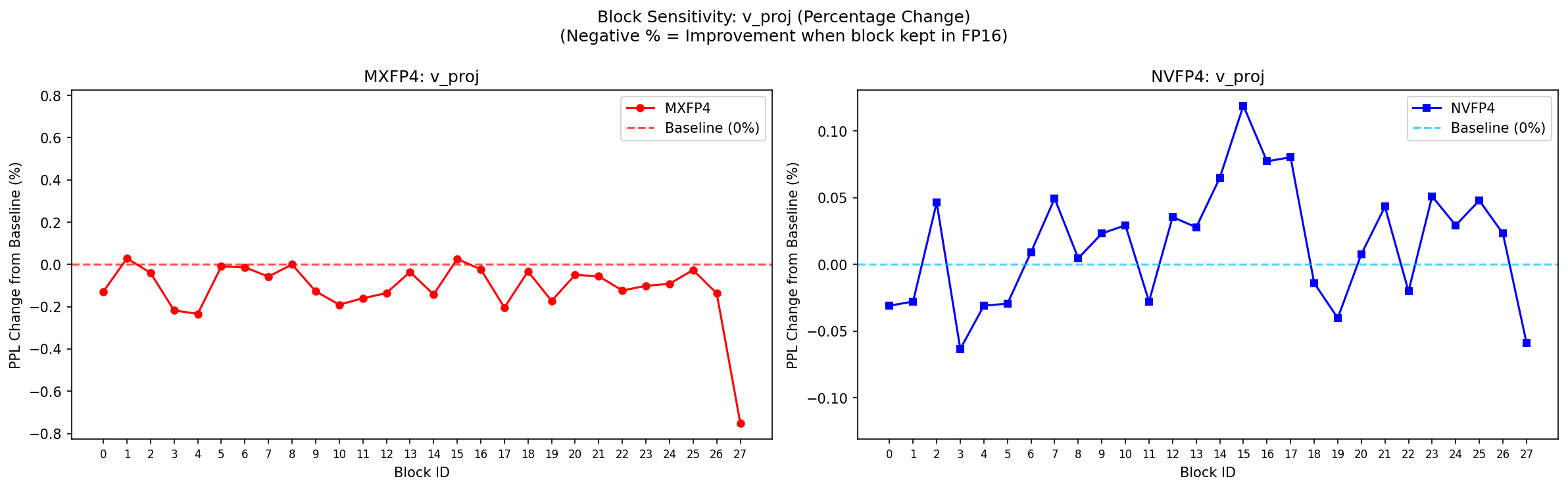}
\end{center}
\vspace{-2.5mm}
\caption{Percentage change from baseline for \texttt{v\_proj} (7B).}
\label{fig:block_v_proj_7b_pct}
\end{figure}

\begin{figure}[H]
\begin{center}
\includegraphics[width=0.85\linewidth]{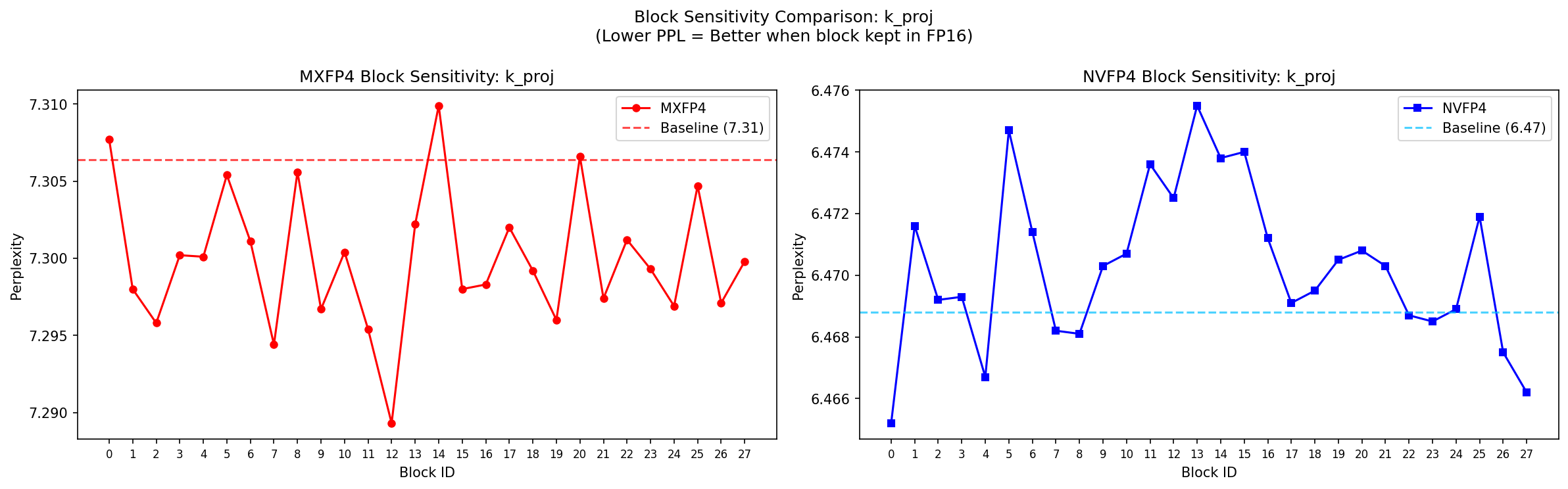}
\end{center}
\vspace{-2.5mm}
\caption{Block sensitivity for \texttt{k\_proj} (7B). Both formats show relatively low and uniform sensitivity across blocks.}
\label{fig:block_k_proj_7b}
\end{figure}

\begin{figure}[H]
\begin{center}
\includegraphics[width=0.85\linewidth]{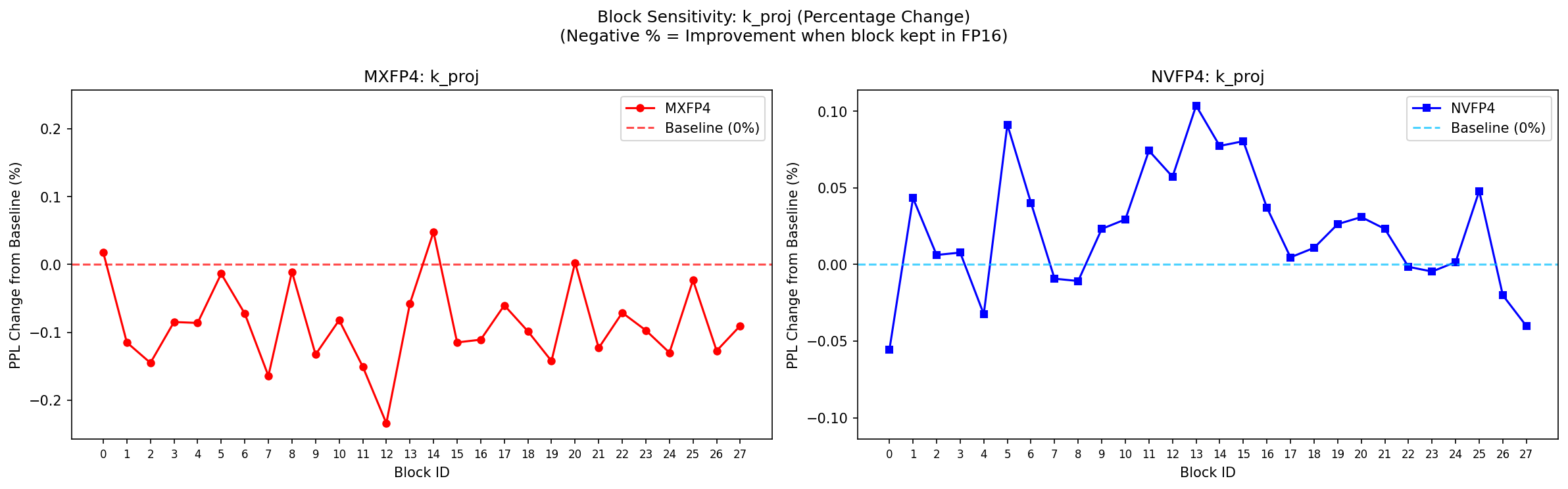}
\end{center}
\vspace{-2.5mm}
\caption{Percentage change from baseline for \texttt{k\_proj} (7B).}
\label{fig:block_k_proj_7b_pct}
\end{figure}

\begin{figure}[H]
\begin{center}
\includegraphics[width=0.85\linewidth]{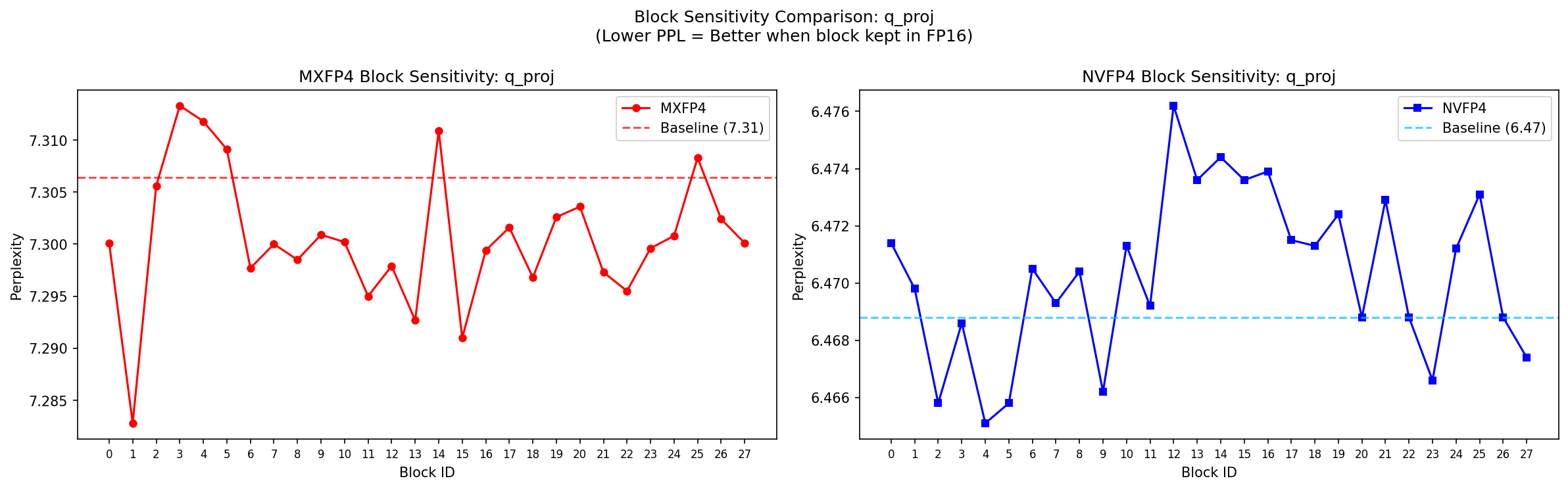}
\end{center}
\vspace{-2.5mm}
\caption{Block sensitivity for \texttt{q\_proj} (7B). MXFP4 shows block 1 as most sensitive, while NVFP4 shows minimal variation across blocks.}
\label{fig:block_q_proj_7b}
\end{figure}

\begin{figure}[H]
\begin{center}
\includegraphics[width=0.85\linewidth]{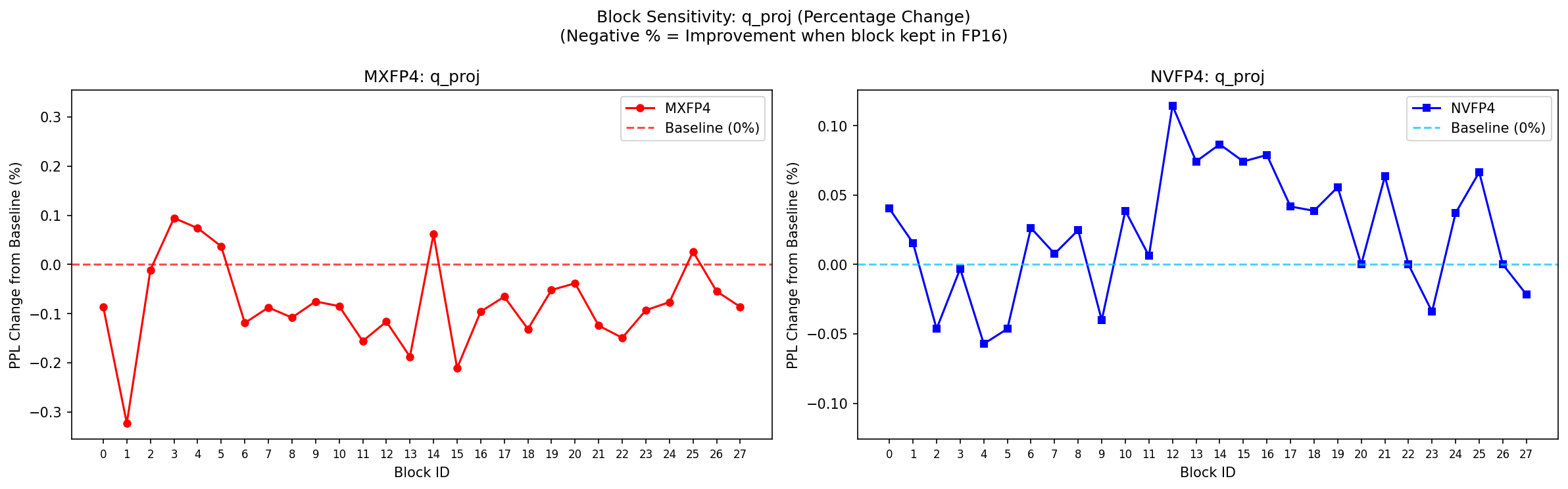}
\end{center}
\vspace{-2.5mm}
\caption{Percentage change from baseline for \texttt{q\_proj} (7B).}
\label{fig:block_q_proj_7b_pct}
\end{figure}

\end{document}